\documentclass[prd,preprint]{revtex4-1}%
\usepackage[dvips]{graphicx}
\usepackage{color}
\usepackage{amsmath}
\usepackage{amsfonts}
\usepackage{amssymb}
\usepackage[colorlinks=true, pdfstartview=FitV, linkcolor=red, citecolor=blue,
urlcolor=blue]{hyperref}%
\newcommand{\be}{\begin{equation}}
\newcommand{\ee}{\end{equation}}
\newcommand{\beq}{\begin{equation}}
\newcommand{\eeq}{\end{equation}}
\newcommand{\bea}{\begin{eqnarray}}
\newcommand{\eea}{\end{eqnarray}}

\graphicspath{{allplotssu2/}}
\begin{document}

\preprint{RBRC-1027} \preprint{BNL-101068-2013-JA}

\title{Matrix model for deconfinement in a $SU(2)$ gauge theory in $2+1$ dimensions}

\author{P. Bicudo}
\affiliation{CFTP, Dep. F\'{\i}sica, 
Instituto Superior T\'ecnico, Universidade T\'ecnica de Lisboa, 
Av. Rovisco Pais, 1049-001 Lisboa, Portugal}

\author{Robert D. Pisarski}
\affiliation{Department of Physics, 
Brookhaven National Laboratory, Upton, NY 11973 \\
and RIKEN/BNL, Brookhaven National Laboratory, Upton, NY 11973}

\author{E. Seel}
\affiliation{Institute of Theoretical Physics, 
J. W. Goethe University, Max-von-Laue Str. 1, D-60438, 
Frankfurt am Main, Germany}

\begin{abstract}
We use matrix models to characterize deconfinement at a nonzero temperature
$T$ for an $SU(2)$ gauge theory in three spacetime dimensions. At one loop
order, the potential for a constant vector potential $A_{0}$ is $\sim T^{3}$
times a trilogarithm function of $A_{0}/T$. In addition, we add various
nonperturbative terms to model deconfinement. The parameters of the model are
adjusted by fitting the lattice results for the pressure. The nonperturbative
terms are dominated by a constant term $\sim T^{2}T_{d}$, where $T_{d}$ is the
temperature for deconfinement. Besides this constant, we add terms which are
nontrivial functions of $A_{0}/T$, both $\sim T^{2}\,T_{d}$ and $\sim
T\,T_{d}^{2}$. There is only a mild sensitivity to the details of these
nonconstant terms. Overall we find a good agreement with the lattice results.
For the pressure, the conformal anomaly, and the Polyakov loop the nonconstant
terms are relevant only in a narrow region below $\sim1.2\,T_{d}.$ We also
compute the 't Hooft loop, and find that the details of the nonconstant terms
enter in a much wider region, up to $\sim4\,T_{d}.$

\end{abstract}
\maketitle

\section{Introduction \label{sec:1}}

Understanding the phase transitions of a non-Abelian gauge theory is of
intrinsic interest, and of relevance to the collisions of heavy ions at
ultrarelativistic energies. Numerical simulations on the lattice provide
detailed results for the pressure and other quantities in equilibrium. This
includes results in the pure gauge theory (without dynamical quarks) for three
colors \cite{Boyd:1996bx, *Umeda:2008bd, *Borsanyi:2012vn}; for the pure
$SU(N)$ theory when $N> 3$ \cite{Lucini:2012gg}, and with dynamical quarks,
Refs. \cite{DeTar:2009ef, *Petreczky:2012rq}.

Besides the theory in four space-time dimensions, it is also useful to
consider gauge theories in three dimensions. For the pure glue theory, the
behavior appears similar in three and four space-time dimensions. There is
confinement at zero temperature, with a linear potential between (external)
quarks in the fundamental representation. This linear potential is
characterized by a string tension, $\sigma$.

At nonzero temperature, numerical simulations on the lattice indicate that for
both theories, there is a deconfining transition at a temperature $T_{d}$. The
results of simulations in three dimensions are given in
\cite{Bialas:2000ev,Bialas:2004gx,Bialas:2008rk,Bialas:2009pt,Bialas:2012qz,Caselle:2011fy,
Caselle:2011mn}.

There are some differences between deconfinement in three and four dimensions.
For example, in an elementary string model \cite{Pisarski:1982cn}, in $d$
space-time dimensions the relationship between the deconfinement temperature
and the string tension is%
\begin{equation}
T_{d}=\sqrt{\frac{3\,\sigma}{\pi(d-2)}}\;. \label{string_tension}%
\end{equation}
The deconfinement temperature is infinite in two dimensions, as then the pure
glue theory is a free field theory (consider, e.g., $A_{0}=0$ gauge). This
ratio decreases as $d$ increases, equal to $T_{d}/\sqrt{\sigma}\approx0.98$ in
three dimensions and $\approx0.67$ in four. These values are in good agreement
with the lattice results of Refs.
\cite{DeTar:2009ef,Petreczky:2012rq,Bialas:2009pt}. For an $SU(N)$ theory, the
order of the transition also changes, as infrared fluctuations in two spacial
dimensions drive the transition{ to }second order even for three colors, where
mean field theory predicts a first-order transition.

We also note that gauge theories in three dimensions may also be relevant for
theories of high temperature superconductivity \cite{Lee:2006zzc}.

For the pure glue theory, the results of lattice simulations are close to the
continuum limit. This is much harder with dynamical quarks, especially those
that are light. Moreover, while numerical simulations can directly compute
many quantities in thermal equilibrium, obtaining results for quantities near
equilibrium is rather more challenging. Such quantities are often of greatest
interest to experiment, such as for transport coefficients like the shear viscosity.

Consequently, it is useful to have approximate models to model the deconfining
transition. One such class of theories are matrix models
\cite{Pisarski:2000eq, Meisinger:2001cq, Meisinger:2001fi, Dumitru:2003hp,
*Oswald:2005vr, Pisarski:2006hz, *Pisarski:2006yk, Hidaka:2008dr,
*Hidaka:2009hs, *Hidaka:2009xh, *Hidaka:2009ma, Dumitru:2010mj,
Dumitru:2012fw, Pisarski:2012bj, Kashiwa:2012wa, Sasaki:2012bi,
*Ruggieri:2012ny, *Diakonov:2012dx, Ogilvie:2012is, Kashiwa:2013rm,
Lin:2013qu}. These involve zero \cite{Meisinger:2001cq, Meisinger:2001fi}, one
\cite{Dumitru:2010mj}, and two \cite{Dumitru:2012fw} parameters, and have been
used to compute various quantities for gauge theories in four dimensions. Such
models dominate for a gauge theory on a femtosphere \cite{Ogilvie:2012is}.

These matrix models are manifestly effective theories. Their virtue is
simplicity. It is known {that in} the pure gauge theory the Polyakov loop
approaches one at infinitely high temperature, and vanishes below $T_{d}$.
This can be modeled by constructing an effective theory for the eigenvalues of
the Wilson line. The relevant variables are $A_{0}/T$, where $A_{0}$ is the
timelike component of the vector potential. One then adds, by hand, terms
which are functions of $A_{0}/T$, to drive the transition to confinement. For
an $SU(N)$ theory, this approach is reasonable at infinite $N$, where this
$A_{0}$ field represents a master field for deconfinement.

The parameters of the matrix models are determined by fitting to the lattice
data for the pressure. Numerical simulations on the lattice gives detailed
data on the pressure as a function of temperature, $p(T)$. It is also useful
to compute other quantities, such as the interaction measure in four
dimensions, $\left[  e(T)-3p(T)\right]  /T^{4}$, where $e(T)$ is the energy
density. This vanishes in the conformal limit, and so naturally characterizes
the deviations from ideality.

In four dimensions, lattice simulations find that, to a good approximation,
the interaction measure, times $T^{2}/T_{d}^{2}$, is constant from
$\sim1.2\,T_{d}$ to $\sim4.0\,T_{d}$ \cite{Boyd:1996bx, Meisinger:2001cq,
Pisarski:2006hz, *Pisarski:2006yk, Umeda:2008bd, Borsanyi:2012vn}. An
approximately constant value of interaction measure, times $T^{2}/T_{d}^{2}$,
implies that the pressure is dominated by a constant term $\sim T_{d}^{2}%
T^{2}$. In the following we refer to terms independent of $A_{0}$ as constant,
and to terms which depend on $A_{0}$ as nonconstant. One finds that when
scaled by the pressure of an ideal gas of gluons, the ratio $p(T)/p_{ideal}%
(T)$ grows sharply for $\sim1.2\,T_{d}<T<$ $\sim4.0\,T_{d}$. This range is
also called the semi-quark gluon plasma (semi-QGP). \ For the pressure the
details of the matrix model matter only in a narrow transition region, from
$T_{d}$ to $\sim1.2\,T_{d}$. In contrast to the pressure, the 't Hooft loop,
for example, is sensitive to the details of the matrix model in a much wider
region, up to $4.0\,T_{d}$ \cite{Dumitru:2010mj, Dumitru:2012fw}.

In this paper we consider a matrix model for an $SU(2)$ gauge theory in three
space-time dimensions. As in four dimensions, we find that the matrix model
works reasonably well even for two colors. The major reason for studying two
colors is technical. After diagonalizing the constant matrix $A_{0}/T$, for
$SU(N)$ the matrix model is a function of the $N-1$ mutually commuting
eigenvalues. For two colors there is only one such eigenvalue, greatly
simplifying the computations.

Broadly, we find that the model in three dimensions looks similar to that in
four dimensions. The interaction measure in three dimensions, $\left[
e(T)-2p(T)\right]  /T^{3}$, times a single power of $T/T_{d}$, is
approximately constant from $\sim1.2\,T_{d}$ to $\sim10\,T_{d}$
\cite{Bialas:2008rk, Caselle:2011mn}. This implies that in this region, the
pressure is dominated by a constant term $\sim T^{2}T_{d}$.

In three dimensions the one- and the two-parameter matrix models are in
reasonable agreement with the lattice results for the pressure. However, near
$T_{d}$ there are significant differences between the matrix model and the
lattice data for the interaction measure. We then introduce a four-parameter
fit which improves the agreement with the lattice data, and reproduces the
correct shape for the interaction measure near $T_{d}$. In this four-parameter
fit, the Polyakov loop deviates from one over a narrow region, up to
$\sim1.2\,T_{d}$. In contrast, for the 't Hooft loop the details of the matrix
model are relevant over a much broader region, up to $\sim4.0\,T_{d}$. The 't
Hooft loop also exhibits only a mild dependence on the details of the
nonconstant terms in the effective Lagrangian.

The outline of the paper is as follows. In Sec. \ref{sec:1} we introduce the
basic concept of the matrix model, and give the motivation to study it in
three space-time dimensions. In Secs. \ref{sec:2} and \ref{sec:3} we construct
the effective potential using the four dimensional case as a guideline: \ In
Sec. \ref{sec:2} we calculate the perturbative potential to one-loop order,
and in Sec. \ref{sec:3} we model the nonperturbative contributions. In Sec.
\ref{sec:4} we present the analytical solution to the effective potential, and
in Sec. \ref{sec:5} we show the numerical fits to the lattice pressure and to
the interaction measure. In Sec. \ref{sec:6} we compute the interface tension
and present the plots for the Polyakov loop and for the 't Hooft loop.
Finally, in Sec. \ref{sec:7} we summarize our results and give an outlook.

\section{Perturbative Potential \label{sec:2}}

In the imaginary-time formalism, the partition function of an $SU(2)$ gauge
theory at a temperature $T$ is%
\begin{equation}
Z=\int DA_{\mu}\exp\left\{  -\frac{1}{4}\int_{0}^{1/T}d\tau\int\,d^{2}%
x\,\operatorname{tr}\,G_{\mu\nu}\,G_{\mu\nu}\right\}  \text{ ,}%
\end{equation}
where $A_{\mu}=iA_{\mu}^{a}\sigma^{a}/2$ is the gauge potential, $\sigma^{a}$
are the Pauli matrices, and $G_{\mu\nu}=\partial_{\mu}A_{\nu}-\partial_{\nu
}A_{\mu}-i\,g\,[A_{\mu},A_{\nu}]$ is the field-strength tensor. In $2+1$
dimensions $A_{\mu}$ and the coupling constant $g$ both have dimensions of
mass$^{1/2}$. Thus, results to one-loop order are proportional to $g^{2}$,
which has the dimensions of mass.

The goal is to construct a model to describe the confinement-deconfinement
phase transition in $SU(2)$. We begin by computing the perturbative potential
in the presence of a constant background field%
\begin{equation}
A_{0}=A_{0}^{cl}+A_{0}^{qu}\;.
\end{equation}
$A_{0}^{cl}$ is a constant classical field%
\begin{equation}
A_{0}^{cl}=\;\frac{\pi\,T\,q}{g}\;\sigma_{3}\text{ ,} \label{a0}%
\end{equation}
where $\sigma_{3}$ is the diagonal $SU(2)$ Pauli matrix,%
\begin{equation}
\sigma_{3}=\left(
\begin{array}
[c]{cc}%
1 & 0\\
0 & -1
\end{array}
\right)  \text{ ,}%
\end{equation}
and $A_{0}^{qu}$ denotes quantum fluctuations.

In this background field the Wilson line is%
\begin{equation}
\mathbf{L}\left(  \vec{x}\right)  =\mathcal{P}\exp\left[  ig\int_{0}%
^{1/T}A_{0}\left(  \vec{x},\tau\right)  d\tau\right]  =\left(
\begin{array}
[c]{cc}%
e^{i\pi q} & 0\\
0 & e^{-i\pi q}%
\end{array}
\right)  \;.
\end{equation}
The eigenvalues of the Wilson line are given by $\mathrm{e}^{\pm i\pi q}$.
They are the basic variables of this model. The relationship between a
background $A_{0}$ field and the eigenvalues of the Wilson line becomes more
complicated at two-loop order and beyond, but this can be ignored to one-loop
order. The Polyakov loop is the trace of the Wilson line%
\begin{equation}
l=\frac{1}{2}\operatorname{tr}\mathbf{L=}\cos\left(  \pi q\right)  \;.
\label{p}%
\end{equation}
Equation (\ref{a0}) is the simplest ansatz which generates a nontrivial
expectation value for the Polyakov loop. Notice, within our model the Polyakov
loop differs from unity only if $q\neq0$\textbf{.}

One perturbative vacuum is given by $A_{0}=q=0$, where $\mathbf{L}=\mathbf{1}$
and $l=1$. The pure gauge $SU(2)$ theory is invariant under global $Z(2)$
gauge rotations. Reflecting this $Z(2)$ symmetry, an equivalent perturbative
vacuum occurs at $q=1$, where $\mathbf{L}=-\mathbf{1}$ and $l=-1$. As a
periodic variable, normally one would expect $q$ to vary from $0\rightarrow2$.
Because of the $Z(2)$ symmetry we can be more restrictive and require $q$ to
lie in the interval from $0\rightarrow1$. If we require $q$ to lie in this
interval, a global $Z(2)$ transformation is given by%
\begin{equation}
q\rightarrow1-q\text{ : \ \ }\mathbf{L\rightarrow}=\left(  -\right)  \left(
\begin{array}
[c]{cc}%
e^{-i\pi q} & 0\\
0 & e^{i\pi q}%
\end{array}
\right)  \text{ ; \ }l\rightarrow-\text{ }l\text{ .}%
\end{equation}
The $Z(2)$ symmetry will become important when we construct the effective
potential, as any possible term will have to be invariant under the
transformation $q\rightarrow1-q$. The confining vacuum is given by the point
halfway between these degenerate vacua,%
\begin{equation}
q_{c}=\frac{1}{2}\;;\;\ \mathbf{L}_{c}=\left(
\begin{array}
[c]{cc}%
i & 0\\
0 & -i
\end{array}
\right)  \text{ ; \ }l_{c}=0\;.
\end{equation}
\label{conf_two} Thus, one can model the transition to deconfinement by
introducing potentials for $q$. It is important to stress that this assumes
that the expectation value of the Polyakov loop is dominated by the classical
configuration of Eq. (\ref{a0}). This is certainly valid at infinite $N$. It
is less obvious that such a master field applies even for two colors.
Nevertheless, one finds that this classical approximation provides a
reasonable ansatz.

Assuming that confinement is dominated by the classical configuration of Eq.
(\ref{a0}) does not provide any understanding of what type of the effective
Lagrangian can produce such a state. This is the principal task of
constructing matrix models for deconfinement. However, there are perturbative
contributions to the free energy in this background field. This has been
computed previously in four dimensions by many authors; see, e.g., Ref.
\cite{Bhattacharya:1990hk, Bhattacharya:1992qb}. In three dimensions it was
computed in Ref. \cite{KorthalsAltes:1996xp}. This classical field is directly
relevant for the computation of the $Z(N)$ interface tension
\cite{Bhattacharya:1990hk, Bhattacharya:1992qb, KorthalsAltes:1996xp}, which
is equivalent to the string tension of the 't Hooft loop
\cite{KorthalsAltes:1999xb, *KorthalsAltes:2000gs}.

To one-loop order the perturbative potential is%
\begin{equation}
V_{pt}(q)=\frac{T}{2\mathcal{V}}\operatorname{tr}\ln\left[  -D^{2}(q)\right]
\text{ }, \label{vpt1}%
\end{equation}
where $\mathcal{V}$ is the two dimensional spacial volume. $D_{\mu}\left(
q\right)  $ denotes the covariant derivative in the adjoint representation, in
the presence of the background $A_{0}$ field of Eq. (\ref{a0})%
\begin{align}
D_{\mu}(q)  &  =\partial_{\mu}-ig\left[  A_{\mu}\mathbf{,}\text{ }\right]
\nonumber\\
&  =\partial_{\mu}-i\pi qT\delta_{\mu,0}\left[  \sigma_{3}\mathbf{,}\text{
}\right]  \text{ .}%
\end{align}
$D^{2}(q)$ is the associated gauge covariant d'Alembertian%
\begin{equation}
D^{2}(q)=\left(  \partial_{0}-i\pi qT\left[  \sigma_{3}\mathbf{,}\text{
}\right]  \right)  ^{2}+\vec{\partial}^{2}\text{ ,}%
\end{equation}
and $\left[  \sigma_{3}\mathbf{,}\text{ }\right]  $ denotes the adjoint
operator%
\begin{equation}
\left[  \sigma_{3},\text{ }\right]  t=\left[  \sigma_{3},\text{ }t\right]
\text{ .}%
\end{equation}
To proceed one needs to introduce a suitable parametrization for the
generators of $SU(2).$ It is useful to choose a ladder basis
\cite{Bhattacharya:1992qb}%
\begin{equation}
t^{+}=\frac{1}{\sqrt{2}}\left(
\begin{array}
[c]{cc}%
0 & 1\\
0 & 0
\end{array}
\right)  \text{ },\text{\ \ }t^{-}=\frac{1}{\sqrt{2}}\left(
\begin{array}
[c]{cc}%
0 & 0\\
1 & 0
\end{array}
\right)  \text{ ,\ \ \ }t_{3}=\frac{1}{2}\left(
\begin{array}
[c]{cc}%
1 & 0\\
0 & -1
\end{array}
\right)  \text{\ ,}%
\end{equation}
where $t_{3}$ is proportional to the diagonal Pauli matrix $\sigma_{3},$ and
$t^{\pm}$ are the off-diagonal step operators. These generators form an
orthogonal set, with the normalization%
\begin{equation}
\operatorname{tr}\left(  t_{3}^{2}\right)  =\frac{1}{2}\text{ , \ }%
\operatorname{tr}\left(  t^{+}t^{-}\right)  =\frac{1}{2}\text{ ,
\ }\operatorname{tr}\left(  t^{+}t^{+}\right)  =\operatorname{tr}\left(
t^{-}t^{-}\right)  =0\text{\ .} \label{norm}%
\end{equation}
The trace in Eq. (\ref{vpt1}) is over all color degrees of freedom. The
diagonal mode $\sim t_{3}$ commutes with the background field. So, the
covariant derivative associated with the diagonal degree of freedom is
independent of $q$:%
\begin{equation}
D_{\mu}\sigma_{3}=\partial_{\mu}\sigma_{3\text{ }},
\end{equation}
and the potential is as in zero background field. The two off-diagonal modes
$\sim$\ $t^{\pm}$ do not commute with $A_{0}^{cl},$%
\begin{equation}
\left[  \sigma_{3}\mathbf{,}\text{ }t^{\pm}\right]  =\pm2t^{\pm}\text{ }.
\end{equation}
They give a nontrivial potential for $q$. The quantum correction enters by
replacing $\partial_{0}$ by $\partial_{0}\pm i2\pi Tq$ in the covariant
derivative%
\begin{equation}
D_{0}t^{\pm}=\left(  \partial_{0}-i\pi qT\left[  \sigma_{3}\mathbf{,}\text{
}\right]  \right)  t^{\pm}=i2\pi T\left(  n\mp q\right)  t^{\pm}.
\end{equation}
In momentum space, the propagators along the off-diagonal degrees of freedom
are as in zero background field, except that the energy $k_{0}$ is shifted to
$k_{0}^{\pm}=i2\pi T\left(  n\pm q\right)  $. As a bosonic field the gluon
must satisfy periodic boundary conditions, which require that $n$ is an
integer, $n=0,\pm1,\pm2\ldots$.

Summing over the diagonal and the off-diagonal modes, the full one-loop result
for the perturbative\ potential in the background field of Eq. (\ref{a0}) is%
\begin{equation}
V_{pt}(q)=\frac{T}{2\mathcal{V}}\left\{  \operatorname{tr}\ln\left(  k_{0}%
^{2}+k^{2}\right)  +\operatorname{tr}\ln\left[  \left(  k_{0}^{+}\right)
^{2}+k^{2}\right]  +\operatorname{tr}\ln\left[  \left(  k_{0}^{-}\right)
^{2}+k^{2}\right]  \right\}  \;.\label{vpt22}%
\end{equation}
The trace over momenta in Eq. (\ref{vpt22}) is evaluated using contour
integration \cite{Gross:1980br},%
\begin{align}
\operatorname{tr}\ln\left[  \left(  k_{0}^{\pm}\right)  ^{2}+k^{2}\right]   &
=2\mathcal{V}\int\frac{d^{2}k}{\left(  2\pi\right)  ^{2}}\ln\left(
1-e^{-\left\vert \mathbf{k}\right\vert /T\pm i2\pi q}\right)  \nonumber\\
&  =-\frac{\mathcal{V}\,}{\pi}\int_{0}^{\infty}dk\;k\sum_{n=1}^{\infty}%
\frac{e^{-nk/T\pm i2\pi qn}}{n}\nonumber\\
&  =-\frac{\mathcal{V}\,T^{2}}{\pi}\sum_{n=1}^{\infty}\frac{e^{\pm i2\pi qn}%
}{n^{3}}\;.\label{free_energy}%
\end{align}
The sum over $n$ converges quickly, and so it can easily be evaluated
numerically \cite{KorthalsAltes:1996xp}. It is also useful to recognize that
this sum can be written in terms of the polylogarithm function,%
\begin{equation}
\mathrm{Li}_{j}(z)=\sum_{n=1}^{\infty}\;\frac{z^{n}}{n^{j}}\;.
\end{equation}
To one-loop order the perturbative potential for $q$ involves the
polylogarithm function of the third kind, which is the trilogarithm function,%
\begin{equation}
V_{pt}(q)=-\frac{T^{3}}{2\pi}\left[  \text{Li}_{3}\left(  e^{i2\pi q}\right)
+\text{Li}_{3}\left(  e^{-i2\pi q}\right)  +\mathrm{Li}_{3}(1)\right]
\;.\label{fqu1}%
\end{equation}
This expression is manifestly symmetric under $Z(2)$ transformations, where
$q\rightarrow1-q$. In Eq. (\ref{fqu1}), the last term, $\mathrm{Li}%
_{3}(1)=\zeta(3)\approx1.202...$, is due to the free energy of the diagonal
mode. In zero field we obtain,%
\begin{equation}
V_{pt}(0)=-\;3\frac{T^{3}}{2\pi}\zeta(3)\;.\label{vpert_zero}%
\end{equation}
In total, this value is minus the pressure for three massless bosons in
$d=2+1$. Note that unlike the four-dimensional case, in three space-time
dimensions there is no factor for the gluon spin. The full one-loop result of
Eq. (\ref{fqu1}) is then the sum of the zero-field contribution in Eq.
(\ref{vpert_zero}) and of the quantum correction%
\begin{equation}
V_{pt}^{qu}(q)=-\frac{T^{3}}{2\pi}\left[  \text{Li}_{3}\left(  e^{i2\pi
q}\right)  +\text{Li}_{3}\left(  e^{-i2\pi q}\right)  -2\mathrm{Li}%
_{3}(1)\right]  \;.\label{vqu}%
\end{equation}


\section{Non-perturbative terms in the effective potential \label{sec:3}}

\subsection{Four dimensions}

Before considering the types of terms which can be added to model
deconfinement, it is instructive to review what happens in four dimensions. In
$d=3+1$ the perturbative term for two colors is given by%
\begin{equation}
V_{pt}^{d=4}(q)=\pi^{2}T^{4}\left[  -\;\frac{1}{15}+\frac{4}{3}\;q^{2}%
(1-q)^{2}\right]  \;. \label{two_color_pot}%
\end{equation}
The term independent of $q$ is the free energy for three gluons, with a factor
of two for the spin. The $q$-dependent term arises from a sum as in three
dimensions, $\sum_{n}\mathrm{e}^{\pm i2\pi q}/n^{4}$. But in $d=3+1$ it
reduces simply to a quartic potential in $q$, $\sim q^{2}(1-q)^{2}.$

There are various nonperturbative terms which one can add to model the
transition to confinement. From the lattice data we know that in four
dimensions the value $(e-3p)/(T^{2}T_{d}^{2})$ is approximately constant from
$1.2\,T_{d}$ to several times $T_{d}$, \cite{Boyd:1996bx, Meisinger:2001cq,
Pisarski:2006hz, *Pisarski:2006yk, Umeda:2008bd, Borsanyi:2012vn}. Taking this
into account, one must certainly add a constant term $\sim T_{d}^{2}\,T^{2}$.
For the pressure, this is the dominant term for temperatures above
$\sim1.2\,T_{d}$.

Similarly, since in three dimensions $(e-2p)/(T^{2}T_{d})$ is constant from
$\sim1.2\,T_{d}$ to $\sim10\,T_{d}$ \cite{Bialas:2008rk, Caselle:2011mn}, one
must also add a constant nonperturbative term $\sim T_{d}T^{2}$ to the
potential for $q$. Referring to such a constant term as nonperturbative, is
somewhat of a misnomer. In three dimensions, the coupling constant squared has
dimensions of mass. Thus at one-loop order, perturbative corrections to the
free energy are $\sim g^{2}T^{2}$, and so automatically proportional to
$T^{3}$. Nevertheless, the results of numerical simulations on the lattice are
still surprising. It is not natural to expect that perturbation theory at
one-loop order is dominant down to temperatures as low as $1.2\,T_{d}$.
Furthermore, the lattice does not indicate the presence of perturbative terms
at two-loop order, which would be $\sim g^{4}T$. Those at three-loop order are
independent of temperature, $\sim g^{6}$. In detail, perturbation theory is
more involved, including logarithms of $g^{2}/T$ \cite{D'Hoker:1980az,
*D'Hoker:1981qp, *D'Hoker:1981us}.

The possible $q$-dependent nonperturbative terms in four dimensions can
certainly include a term like the perturbative potential $\sim q^{2}(1-q)^{2}%
$. In addition, a term linear in $q$ is added. To be consistent with the
$Z(2)$ symmetry, the linear term must be $\sim q(1-q).$ The need for the
linear term can be argued on two grounds. One argument is the following: When
$q$ develops an expectation value, the deconfining phase is in an adjoint
Higgs phase. While there is no gauge-invariant order parameter for an adjoint
Higgs phase, there can still be a first-order transition from a truly
perturbative phase, where $\langle q\rangle=0$, to one where $\langle
q\rangle\neq0$. This would be a second phase transition, at a temperature
higher than $T_{d}$. Though it is possible, the lattice finds no evidence of
such a second phase transition. A term linear in $q$ will give an expectation
value for $q$ at any temperature, obviating the possibility of such a second
phase transition. Another explanation was first discussed by Meisinger and
Ogilvie \cite{Meisinger:2001fi}: \ If one assumes that the gluons develop a
mass, then expanding in the mass squared to leading order, the one-loop
determinant in a background $A_{0}^{cl}$ field is%
\begin{equation}
\frac{T}{\mathcal{V}}\mathrm{tr}\ln\left(  -D_{cl}^{2}+m^{2}\right)  \sim
m^{2}\frac{T}{\mathcal{V}}\mathrm{tr}\left(  \frac{1}{-D_{cl}^{2}}\right)  \;.
\label{linear_4D}%
\end{equation}
In Sec. \ref{sec:vander} we show explicitly how this determinant generates a
term linear in $q$. The origin of this mass term will not be discussed here.
The point is that since the determinant is gauge invariant, the result in Eq.
(\ref{linear_4D}) is gauge invariant as well. Such a term arises naturally in
expanding about the supersymmetric limit. Then $m$ is the mass of an adjoint
fermion, and Eq. (\ref{linear_4D}) is the leading term in an expansion about a
small mass; see \cite{Poppitz:2012nz,Poppitz:2012sw}.

Altogether the possible nonperturbative potential one can construct in four
dimensions is%
\begin{equation}
V_{npt}^{d=4}(q)=-\;T^{2}\,T_{d}^{2}\left[  \frac{1}{5}C_{1}q\left(
1-q\right)  +C_{2}q^{2}\left(  1-q\right)  ^{2}-C_{3}\right]  -B\;.
\label{npt_4}%
\end{equation}
The constant term $\sim C_{3}T^{2}T_{d}^{2}$ is required by the lattice data
for the pressure. It is the dominant term above $\sim1.2\,T_{d}.$
\cite{Boyd:1996bx, Meisinger:2001cq, Pisarski:2006hz, *Pisarski:2006yk,
Umeda:2008bd, Borsanyi:2012vn}. The term $\sim C_{1}$ is required to avoid an
adjoint Higgs phase. This term is also generated by expanding the one-loop
determinant in the mass squared to leading order, Eq. (\ref{linear_4D}), with
$m\sim T_{d}$. Since there is a perturbative term $\sim q^{2}(1-q)^{2}$ in Eq.
(\ref{two_color_pot}), presumably it can also arise in the nonperturbative
potential. It is natural to assume that the temperature dependence of these
nonperturbative terms is $\sim T^{2}T_{d}^{2}$, although this is manifestly an
assumption. Lastly, one can add a term like an MIT bag constant, $B$. This is
the most general model studied to date.

Equation (\ref{npt_4}) involves four-parameters, $C_{1}$, $C_{2}$, $C_{3}$,
and $B.$ Introducing two conditions, one gets a model with only two
independent parameters. The first condition is that the transition occurs at
$T_{d}$, and not at another temperature. The second condition is that the
pressure vanishes at $T_{d}$. The second condition is motivated by large-$N$
arguments, where the pressure is $\sim N^{2}$ in the deconfined phase, and
$\sim1$ in the confined phase. However, especially for two colors, this is a
rather drastic approximation. Instead, one should add an effective theory for
the confined phase, and ensure that the pressures match at $T_{d}$. To date
this has not been done. Consequently, it is not surprising that one finds
unphysical features close to $T_{d}$, within $1\%$ of the transition, such as
a negative pressure \cite{Dumitru:2012fw}. One finds similar unphysical
behavior in three dimensions. But as in four dimensions, we shall view this
purely as a consequence of not matching to a physical equation of state in the
confined phase. We discuss this further when we turn to the results of the
matrix models.

\subsection{Nonperturbative terms in three dimensions}

\subsubsection{Linear terms}

Using the four-dimensional case as a guideline, we add the following terms to
the nonperturbative potential: \ First, we need a constant term $\sim
T^{2}T_{d}$, to ensure that $(e-2p)/(T^{2}T_{d})$ is approximately constant
\cite{Bialas:2008rk, Caselle:2011mn}. Second, it is natural to include a term
similar to that generated in perturbation theory, Eq. ( \ref{vqu}). Lastly, to
avoid a transition to an adjoint Higgs phase above $T_{d}$, one adds a term
linear in $q$ for small $q$, $\sim q(1-q)$. We can also write the linear term
in a more general way by adding a factor plus a constant%
\begin{equation}
bq(1-q)+d\text{ ,} \label{lin}%
\end{equation}
which preserves all the required features and the $Z(2)$ symmetry. Altogether
the nonperturbative potential for $SU(2)$ is%
\begin{align}
V_{npt}^{A}(q)  &  =-T^{2}T_{d}C_{1}\left[  bq(1-q)+d\right]  +T^{2}%
T\,_{d}C_{3}\frac{3\,\zeta\left(  3\right)  }{2\pi}\nonumber\\
&  +T^{2}T_{d}\frac{C_{2}}{2\pi}\left[  \text{Li}_{3}\left(  e^{i2\pi
q}\right)  +\text{Li}_{3}\left(  e^{-i2\pi q}\right)  -2\zeta\left(  3\right)
\right]  \;. \label{npt_A}%
\end{align}
So far we have assumed that all nonperturbative terms are proportional to
$T^{2}T_{d}$. This is necessary for the constant term $\sim C_{3}$, but there
is no such restriction for the $q$-dependent terms. The possibility of a
different temperature dependence for the term $\sim C_{1}$ will be discussed later.


\subsubsection{Vandermonde determinant \label{sec:vander}}

Besides the linear term $\sim q(1-q),$ there is another possibility to
construct a nonperturbative term which is linear in $q$ for small $q:$ \ As in
four dimensions, we consider the expansion of the one-loop determinant to
leading order in a mass parameter%
\begin{equation}
\frac{T}{\mathcal{V}}\mathrm{tr}\ln\left(  -D_{cl}^{2}+m^{2}\right)  \sim
m^{2}\frac{T}{\mathcal{V}}\mathrm{tr}\left(  \frac{1}{-D_{cl}^{2}}\right)
\text{ .} \label{linear_3D1}%
\end{equation}
The simplest way is to follow the computation for zero mass in Eq.
(\ref{free_energy}),%
\begin{align}
\operatorname{tr}\ln\left[  \left(  k_{0}^{\pm}\right)  ^{2}+k^{2}%
+m^{2}\right]   &  =2\mathcal{V}\,\,\int\frac{d^{2}k}{\left(  2\pi\right)
^{2}}\ln\left(  1-e^{-E(k)/T\pm i2\pi q}\right) \nonumber\\
&  =-\frac{\mathcal{V}\,}{\pi}\int_{0}^{\infty}dk\text{ }k\sum_{n=1}^{\infty
}\frac{e^{-n\,E(k)/T\,\pm i\,2\pi q\,n}}{n}\text{ },
\end{align}
where $E(k)=\sqrt{k^{2}+m^{2}}$ is the energy. Now it is easy to compute the
derivative with respect to the mass, and then consider the limit
$m\rightarrow0$%
\begin{align}
\frac{d}{dm^{2}}\operatorname{tr}\ln\left[  \left(  k_{0}^{\pm}\right)
^{2}+k^{2}+m^{2}\right]  _{m^{2}=0}  &  =\frac{\mathcal{V}\,}{2T\pi}\int
_{0}^{\infty}dk\sum_{n=1}^{\infty}e^{-n\,k/T\,\pm\,i2\pi qn}\nonumber\\
&  =\frac{\mathcal{V}\,}{2\pi}\sum_{n=1}^{\infty}\frac{e^{\,\pm i2\pi qn}}%
{n}\nonumber\\
&  =\frac{\mathcal{V}\,}{2\pi}\mathrm{Li}_{1}(\mathrm{e}^{\pm i2\pi q})\text{
.}%
\end{align}
This is a polylogarithm function of the first kind, which can be written as
$\mathrm{Li}_{1}(z)=-\ln(1-z).$ Including both, the contributions of
$k_{0}^{+}$ and $k_{0}^{-}$ gives a result which is automatically real,%
\begin{equation}
\mathrm{tr}\left(  \frac{1}{-D_{cl}^{2}}\right)  =\frac{\mathcal{V}\,}{\pi
}\sum_{n=1}^{\infty}\frac{\mathrm{cos}\left(  2\pi qn\right)  }{n}\;.
\end{equation}
In all we obtain%
\begin{align}
\frac{T}{\mathcal{V}}\mathrm{tr}\left(  \frac{1}{-D_{cl}^{2}}\right)   &
=\frac{T\,}{2\pi}\left[  \mathrm{Li}_{1}(\mathrm{e}^{i2\pi q})+\mathrm{Li}%
_{1}(\mathrm{e}^{-i2\pi q})\right] \nonumber\\
&  =-\frac{T\,}{2\pi}\left\{  \ln\left[  1-\exp\left(  2i\pi q\right)
\right]  +\ln\left[  1-\exp\left(  -2i\pi q\right)  \right]  \right\}
\nonumber\\
&  =-\frac{T\,}{\pi}\ln\left[  2\sin\left(  \pi q\right)  \right]  \;.
\label{vdm1}%
\end{align}

Unlike the linear term, which is $\sim T^{2},$ the term in Eq. (\ref{vdm1}) is
proportional to $T$. This is expected, since it enters times the square of a
mass parameter, Eq. (\ref{linear_3D1}). On the other hand, it is surprising
that this term is identical to the Vandermonde determinant, which enters so
often in matrix models. For a femtosphere, or other small systems, it is
natural that the Vandermonde determinant enters, and dominates
\cite{Ogilvie:2012is}. In large volume, however, it is proportional to
$\delta^{d}(0)$, where $d$ is the dimension of space-time. This is in turn
proportional to $\Lambda^{d}$, where $\Lambda$ is some ultraviolet cutoff,
which vanishes when applying dimensional regularization. Such a
regularization-dependent term is not expected to contribute in the limit of
infinite spatial volume. Thus it is surprising to find that it enters in a
mass expansion in three dimensions. Remarkably, while the Vandermonde term
arises on a femtosphere \cite{Ogilvie:2012is}, it does not arise in a mass
expansion in four dimensions, Eq. (\ref{linear_4D}). A term such as Eq.
(\ref{vdm1}) will ensure that the condensate for $q$ is always nonzero.

It is interesting to mention that performing a mass expansion is just one
possibility to derive the Vandermonde term from the perturbative one-loop
result. A similar Vandermonde term can also be found at the two-loop order in
the perturbative expansion \cite{Bhattacharya:1992qb,
Belyaev:1989bj,Enqvist:1990ae}. An equivalent way to determine the
$q$-dependence of this nonperturbative term is to consider the second
derivative of the perturbative trilogarithm function with respect to $q,$%
\begin{align}
\frac{d^{2}}{dq^{2}}\left\{  {\frac{-1}{2\pi}}\left[  \text{Li}_{3}\left(
e^{i2\pi q}\right)  +\text{Li}_{3}\left(  e^{-i2\pi q}\right)  -2\mathrm{Li}%
_{3}(1)\right]  \right\}   &  =2\pi\left[  \mathrm{Li}_{1}(\mathrm{e}^{i2\pi
q})+\mathrm{Li}_{1}(\mathrm{e}^{-i2\pi q})\right] \nonumber\\
&  =-4\pi\ln\left[  2\sin(\pi q)\right]  \text{ .} \label{vdm}%
\end{align}
Notice, by expanding this expression around $q=1/2,$ and keeping only terms to
order $q^{2}$ we also recover the linear term introduced in Eq. (\ref{lin}),
\begin{equation}
-4\pi\ln\left[  2\sin(\pi q)\right]  =-2\pi\left[  2\ln2-\pi^{2}\left(
q-\frac{1}{2}\right)  ^{2}\right]  +O\left[  \left(  q-\frac{1}{2}\right)
^{4}\right]  \ , \label{vlin}%
\end{equation}
with $b=2\pi^{3}$ and $d=4\pi\ln2-\pi^{3}/2$. Strictly speaking, the
Vandermonde term exhibits a divergence at $q=0$. But, as we will see later,
this divergence does not pose any problem for the present study. This is
because all thermodynamical quantities in this work are computed at the
minimum of the effective potential, where the condensate for $q$ effectively
vanishes at high temperatures, but it is never exactly zero. The linear term
in Eq. (\ref{vlin}) can also be seen as a regularized version of the
Vandermonde term in Eq. (\ref{vdm}).

Replacing the linear term in Eq. (\ref{npt_A}) by the Vandermonde term derived
in Eq. (\ref{vdm}), the nonperturbative potential can be written as%
\begin{align}
V_{npt}^{B}(q)  &  =-\;{C}_{1}\,T^{3-\delta}\,{T_{d}}^{\delta} 4 \pi\ln\left[
2\sin(\pi q)\right]  +C_{3}T^{2}T_{d}\,\frac{3\,\zeta\left(  3\right)  }{2\pi
}\nonumber\\
&  +T^{2}T_{d}\frac{C_{2}}{2\pi}\left[  \text{Li}_{3}\left(  e^{i2\pi
q}\right)  +\text{Li}_{3}\left(  e^{-i2\pi q}\right)  -2\,\zeta\left(
3\right)  \right]  \;, \label{npt_B}%
\end{align}
where $\delta=1,2$ denotes two possible temperature dependences. The value
$\delta=2$ is suggested by the mass expansion. But it is also reasonable to
try $\delta=1,$ which gives the same temperature dependence as for the other
two nonperturbative terms.


\section{Analytical Solution \label{sec:4}}

In this Section the analytical solution to the effective potential is
presented. We discuss how to determine the parameters of the model utilizing
the conditions at $T_{d}$. Further, we explain how to obtain the minimum of
the effective potential, and give the analytical expressions for the pressure
and for the interaction measure. The effective potential is constructed as the
sum of the perturbative result to one-loop order plus the nonperturbative
contributions. Then we can compute the pressure as a function of the
temperature%
\begin{equation}
p\left(  T\right)  =-V_{eff}\left[  q_{min}\left(  T\right)  \right]  \text{
,} \label{pressure}%
\end{equation}
where $q_{min}\left(  T\right)  $ is the minimum of the effective
potential.\ Using the first principle of thermodynamics, we can also calculate
the energy density $e$, and the interaction measure $\Delta$%
\begin{equation}
\text{\ }e\left(  T\right)  =T\frac{dp}{dT}-p\left(  T\right)  \text{ ,
\ \ }\Delta=e\left(  T\right)  -2p\left(  T\right)  \text{ .}%
\end{equation}


\subsection{Linear potential \label{sec:analyticallinear}}

\begin{figure*}[t]
\begin{center}
\includegraphics[
width=0.45\textwidth
]{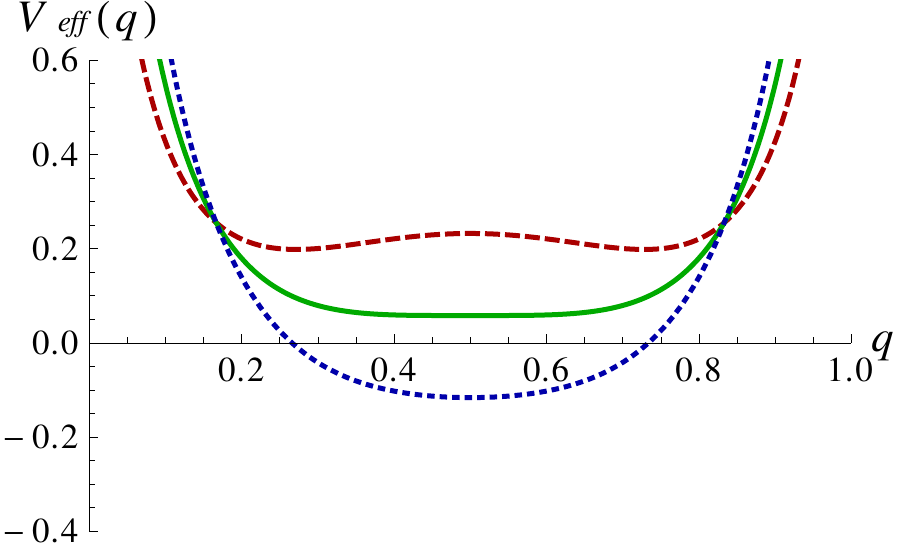} \includegraphics[
width=0.45\textwidth
]{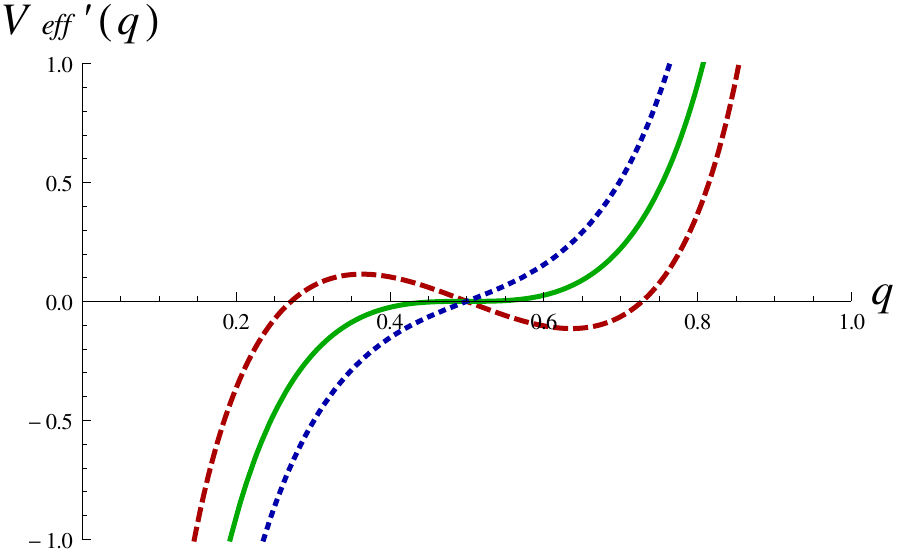}
\end{center}
\caption{ The effective potential, $V_{eff}(q)$ (left panel), and its first
derivative, $V\prime_{eff}(q)$ (right panel), as a function of $q$. We present
the plots for three different values of $a:$ $a<a_{d}$ (dashed) represents the
semi-QGP, at $a=a_{d}$ (solid) the phase transition to confinement takes
place, and for $a>a_{d}$ (dotted) the system is in the confined phase.}%
\label{veffq}%
\end{figure*}

First we discuss the case where the linear term of Eq. (\ref{vlin}) is
used.\ The effective potential is then%
\begin{equation}
V_{eff}=V_{pt}+V_{npt}^{A}\text{ ,}%
\end{equation}
where $V_{pt}$ denotes the perturbative one-loop result of Eq. (\ref{fqu1}),
and $V_{npt}^{A}(q)$ the nonperturbative contributions of Eq. (\ref{npt_A}).
In the following discussion it is useful to rewrite $V_{eff}$ as%
\begin{align}
V_{eff}  &  =-3\frac{\zeta\left(  3\right)  }{2\pi}T^{3}\left(  1-\frac{T_{d}%
}{T}C_{3}\right) \nonumber\\
&  +T^{3}\left(  1-\frac{T_{d}}{T}C_{2}\right)  \left\{  \text{L}\left(
q\right)  -2\pi\left[  2\ln2-\pi^{2}\left(  q-\frac{1}{2}\right)  ^{2}\right]
a(T)\right\}  \text{ ,} \label{veff}%
\end{align}
where we introduce the notation%
\begin{align}
\text{L}\left(  q\right)   &  =-\frac{1}{2\pi}\left[  \text{Li}_{3}\left(
e^{i2\pi q}\right)  +\text{Li}_{3}\left(  e^{-i2\pi q}\right)  -2\zeta\left(
3\right)  \right]  \text{ ,}\nonumber\\
a(T)  &  =\frac{\frac{T_{d}}{T}C_{1}}{\left(  1-\frac{T_{d}}{T}C_{2}\right)
}\text{ .} \label{forveff}%
\end{align}
This effective potential\ exhibits a second-order phase transition, see Fig.
\ref{veffq}. $V_{eff}(q)$ has the shape of a double-well\ potential symmetric
to the confined vacuum $q_{c}=1/2.$ Depending on the value of $a,$ one can
describe the transition from deconfinement to confinement: \ At $a=0,$ the
minima of the effective potential are given by the perturbative vacua at $q=0$
and $q=1.$ This is the region of complete QGP. For $0<a<a_{d}$ the system is
in the semi-QGP phase, and the distance between the confined vacuum and the
two degenerate minima starts decreasing. At $a_{d}=0.070230$ the transition to
confinement takes place, and for $a\geqq a_{d}$ there is just one minimum
which is given by the confined vacuum at $q=1/2$.

\begin{figure*}[t]
\begin{center}
\includegraphics[
width=0.45\textwidth
]{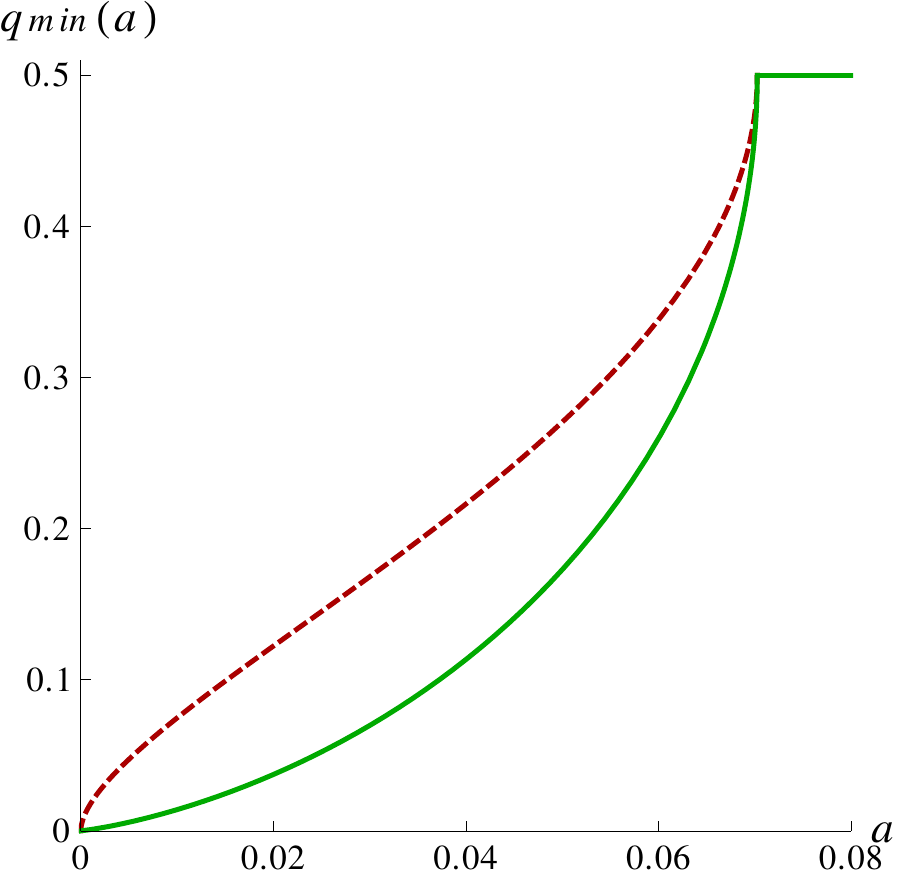} \includegraphics[
width=0.45\textwidth
]{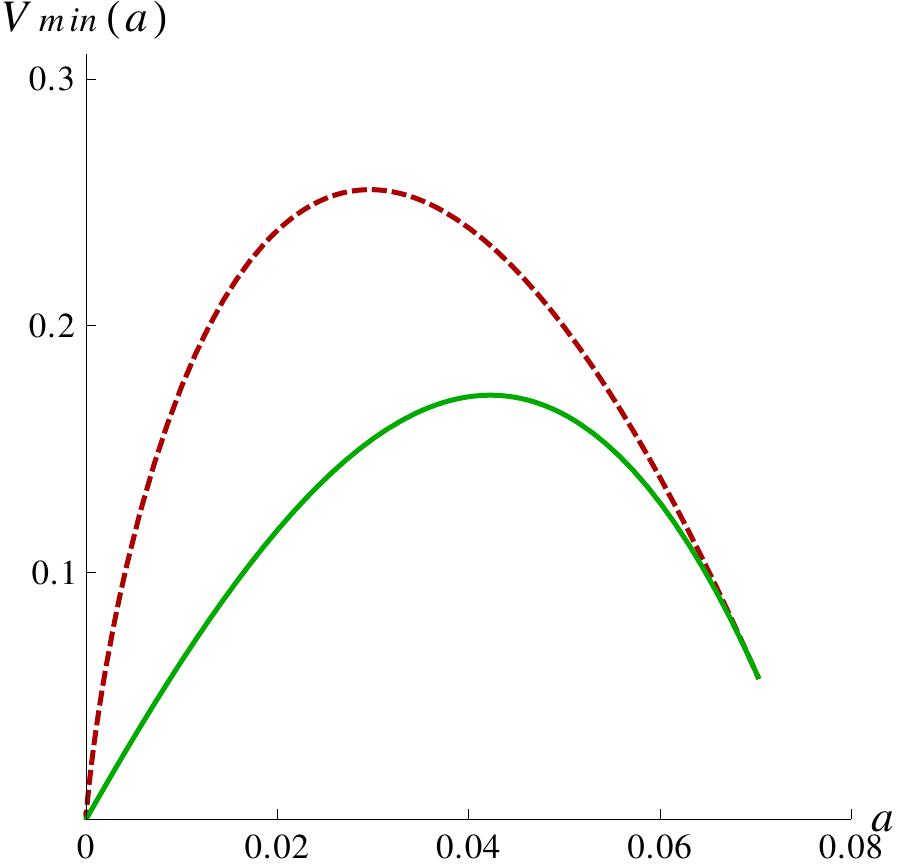}
\end{center}
\caption{Left panel: the minimum of the effective potential as a function of
$a,$ $q_{min}(a),$ using the linear term (solid), and the Vandermonde term
(dashed). Right panel: the potential at the minimum as a function of $a,$
$V_{min}(a)$. Our fits to $q_{min}(a)$ and $V_{min}(a)$ are essentially
identical to the exact numerical solutions, since we work with a high
precision.}%
\label{qmin}%
\end{figure*}

\subsubsection{Fixing the parameter at $T_{d}$}

Apart from $T_{d}$, the effective potential of Eq. (\ref{veff}) involves three
parameters $C_{1},\,C_{2},\,C_{3}$ which are determined from the lattice
measurements of the pressure in the deconfined phase. First, we impose that
the transition occurs at $T_{d}$. This implies that $a(T_{d})=a_{d}$ $:$%
\begin{equation}
a_{d}=\frac{C_{1}}{1-C_{2}}\text{ ,} \label{c1}%
\end{equation}
which gives $C_{2}$ as a function of $C_{1}$. We further require that the
pressure is zero at $T_{d}$ , which allows to determine $C_{3}$%
\begin{equation}
\text{ }C_{3}=1-\frac{C_{1}\left[  \text{L}\left(  0.5\right)  -a_{d}\text{
}8\pi^{2}\ln2\right]  }{3\frac{\zeta\left(  3\right)  }{2\pi}a_{d}}\text{ .}
\label{c3}%
\end{equation}
Due to the two conditions, there is only one free parameter left, say $C_{1}$.
This single parameter is utilized to fit the lattice pressure and the
interaction measure.

\subsubsection{ The minimum of the effective potential}

The main numerical problem to compute the pressure $p(T)$ resides in finding
the minimum of the effective potential as a function of $q$ at $T\geq T_{d}$.
This defines a function $q_{min}(T)$.

For mathematical clarity, it is convenient to denote the $q$-dependent part of
the effective potential in Eq. (\ref{veff}) as $V(q,a),$%
\begin{equation}
V(q,a)=\text{L}\left(  q\right)  -2\pi\left[  2\ln2-\pi^{2}\left(  q-\frac
{1}{2}\right)  ^{2}\right]  a(T)\ .
\end{equation}
The minimum is found by solving numerically the equation
\begin{equation}
{\frac{\partial V(q,a)}{\partial q}\mid}_{q=q_{min}}=0\ , \label{dervtot}%
\end{equation}
for different values of $a,$ in the range $0\leq a\leq a_{d}.$ This gives the
minimum of the effective potential as a function $a,$ $q_{min}(a).$ We use
this solution to obtain an expression for the potential at the minimum, which
depends only on $a$
\begin{equation}
V_{min}(a)=V\left[  q_{min}(a),a\right]  \text{ .}%
\end{equation}
In principle, one needs to solve Eq. (\ref{dervtot}) for every value $0\leq
a\leq a_{d}$ we want to study. However, it is more convenient to find a good
ansatz for $q_{min}(a)$ and for $V_{min}(a)$. Then, it is straightforward to
determine the temperature-dependent minimum by utilizing the definition for
$a(T)$ in Eq. (\ref{forveff})%
\begin{align}
q_{min}(T)  &  =q_{min}\left[  a\left(  T\right)  \right]  =q_{min}\left[
\frac{\frac{T_{d}}{T}C_{1}}{\left(  1-\frac{T_{d}}{T}C_{2}\right)  }\right]
\ ,\nonumber\\
V_{min}(T)  &  =V_{min}\left[  a\left(  T\right)  \right]  =V_{min}\left[
\frac{\frac{T_{d}}{T}C_{1}}{\left(  1-\frac{T_{d}}{T}C_{2}\right)  }\right]
\ . \label{qminT}%
\end{align}
To solve Eq. (\ref{dervtot}) we apply the numerical bisection method. Then we
fit the numerical solutions for $q_{min}(a)$ and $V_{min}(a)$ with high
precision. As an ansatz for the fits we use simple linear expansion in
rational powers of $a,$ and in powers of $a_{d}-a$. The absolute deviation
between the numerical solution and our ansatz for $V_{min}(a)$ is of the order
of $10^{-7}$. It is important to work with good accuracy, because the error
bars of the lattice data for the pressure $p/(3T^{3})$ are small, $10^{-5}$.
In Fig. (\ref{qmin}), we plot the solutions for $q_{min}(a)$ and for
$V_{min}(a).$ Since we use a very high precision, {the curves of our
Ans\"{a}tze coincide with the curves} of the exact numerical solutions.


\subsubsection{Analytical expressions for the pressure and for the interaction
measure}

The pressure as a function of $T$ is obtained by plugging the solution
$q_{min}\left(  T\right)  $ of Eq. (\ref{qminT}) into the equation for the
effective potential of Eq. (\ref{veff})%
\begin{align}
&  {\frac{p}{3T^{3}}}=\left(  1-\frac{T_{d}}{T}C_{3}\right)  \frac
{\zeta\left(  3\right)  }{2\pi}+\frac{2 \pi T_{d}}{3T}C_{1}\left\{  2\ln
2-\pi^{2}\left[  q_{min}(T)-\frac{1}{2}\right]  ^{2}\right\}  \text{
}\label{pressureover3t3}\\
&  \ +\frac{\left(  1-\frac{T_{d}}{T}C_{2}\right)  }{6\pi}\left\{
\text{Li}_{3}\left[  e^{i2\pi q_{min}(T)}\right]  +\text{Li}_{3}\left[
e^{-i2\pi q_{min}(T)}\right]  -2\zeta\left(  3\right)  \right\}  \text{
.}\nonumber
\end{align}
Another possibility to compute the pressure is to directly use the ansatz for
$V_{min}(T)$ {depicted in Fig. (\ref{qmin}) and in Eq. (\ref{qminT}),
\begin{equation}
{\frac{p}{3T^{3}}}=\left(  1-\frac{T_{d}}{T}C_{3}\right)  \frac{\zeta\left(
3\right)  }{2\pi}-\left(  1-\frac{T_{d}}{T}C_{2}\right)  {\frac{\ V_{min}%
(T)}{3}}\text{ }, \label{presurevmin}%
\end{equation}
} which greatly simplifies the numerics.

Differentiating the pressure of Eq. (\ref{pressureover3t3}) with respect to
$T,$\ gives the interaction measure%
\begin{equation}
{\frac{\Delta}{3T^{3}}}=T{\frac{d}{dT}}\left(  \frac{p}{3T^{3}}\right)  \ .
\label{traceover3t3}%
\end{equation}

\begin{figure*}[t]
\begin{center}
\includegraphics[
width=0.90\textwidth]{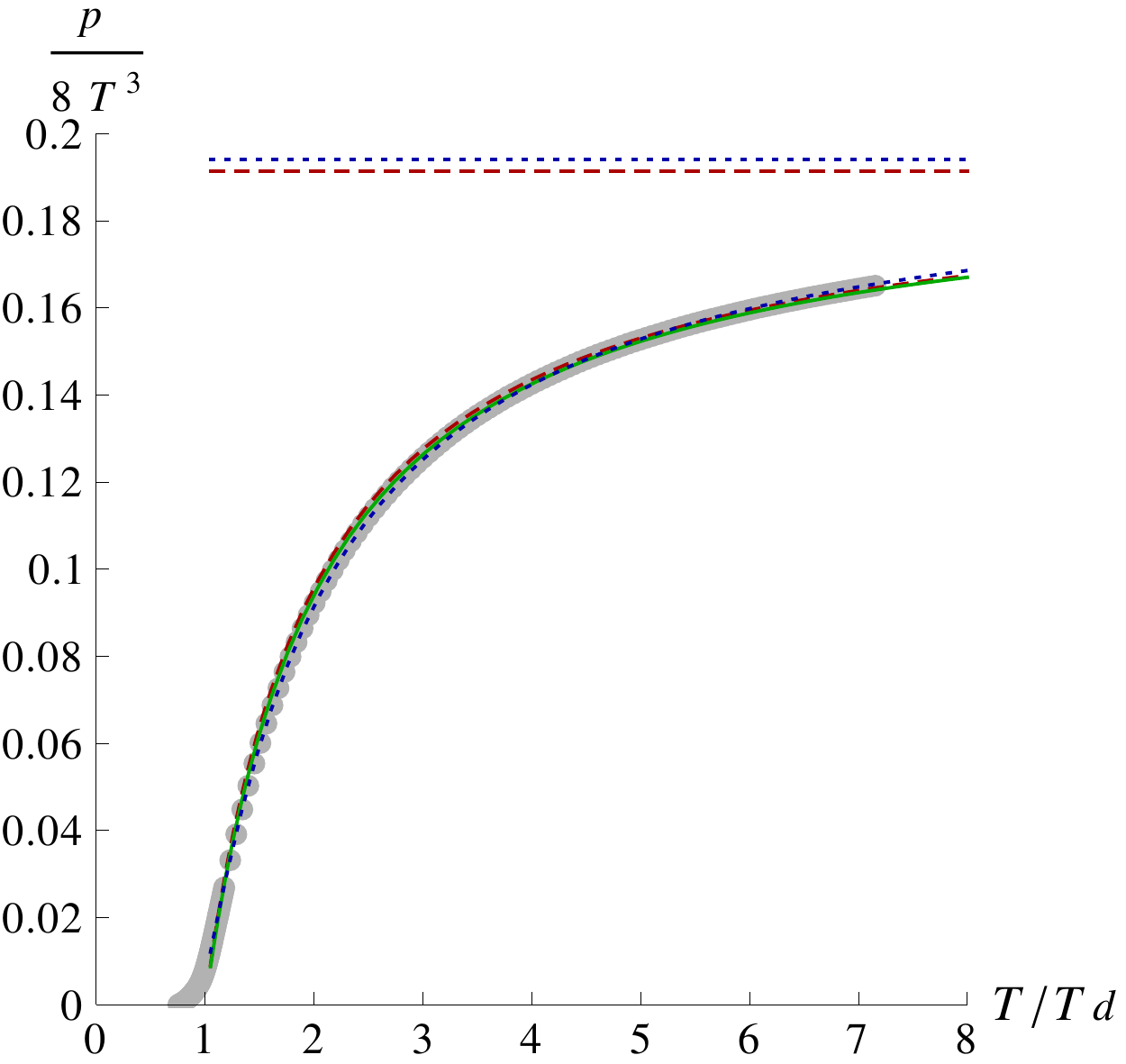}
\end{center}
\caption{Lattice data, as well as the numerical fits to the pressure
$p/3T^{3}$ using the linear term. We present the curves in the one-parameter
model (dashed), the two-parameter model (solid), and in the four-parameter fit
(dotted). The horizontal lines represents the perturbative constant $c,$ which
corresponds to the perturbative limit of the pressure. In the one- and in the
two-parameter models $c=\zeta\left(  3\right)  /2\pi$ (solid), and in the
four-parameter fit it is shifted by $\sim0.5\%.$}%
\label{plin}%
\end{figure*}

\begin{figure*}[t]
\begin{center}
\includegraphics[width=0.90\textwidth]{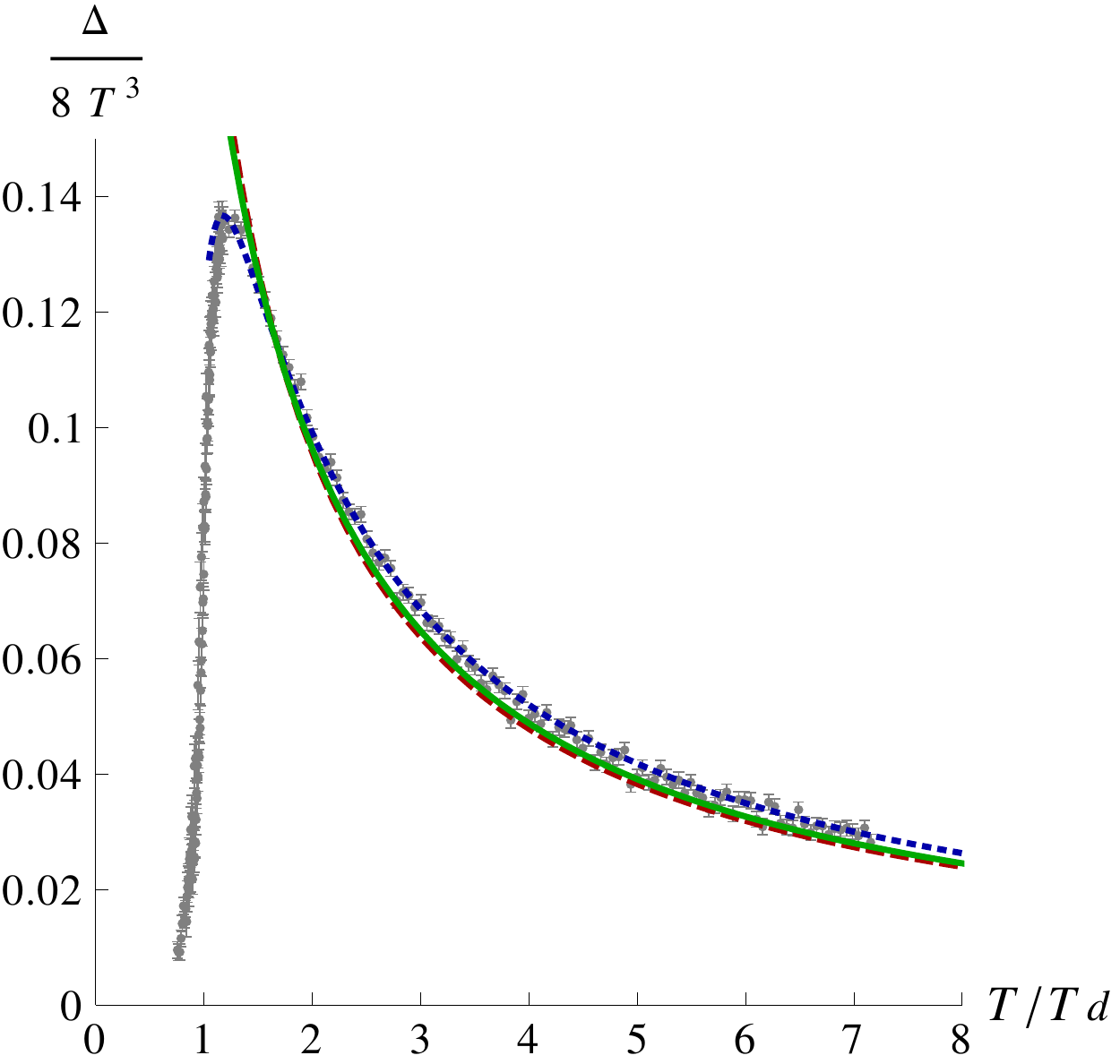}
\end{center}
\caption{Lattice data for the interaction measure $\Delta/3T^{3}$ in
comparison with the results for the linear term. We present the curves in the
one-parameter model (dashed), the two-parameter model (solid), and in the
four-parameter fit (dotted). }%
\label{trlin}%
\end{figure*}

\begin{figure*}[t]
\begin{center}
\includegraphics[width=0.90\textwidth]{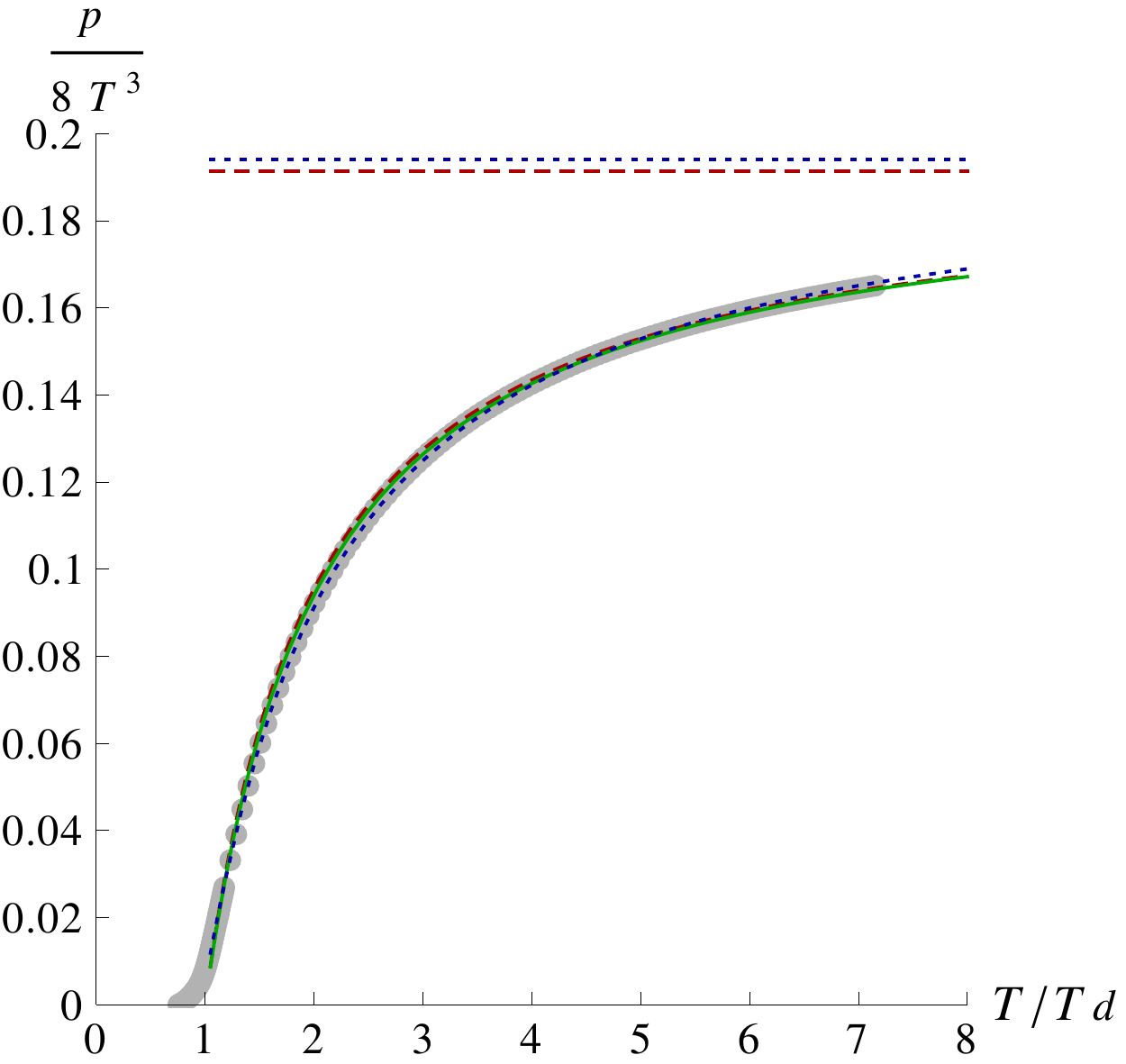}
\end{center}
\caption{Numerical fits to the lattice pressure $p/3T^{3}$ using the
Vandermonde term $\sim T^{2}$. We show the curves in the one-parameter model
(dashed), the two-parameter model (solid), and in the four-parameter fit
(dotted). The horizontal lines represents the pressure in the perturbative
limit.}%
\label{pvdm}%
\end{figure*}

\begin{figure*}[t]
\begin{center}
\includegraphics[width=0.90\textwidth]{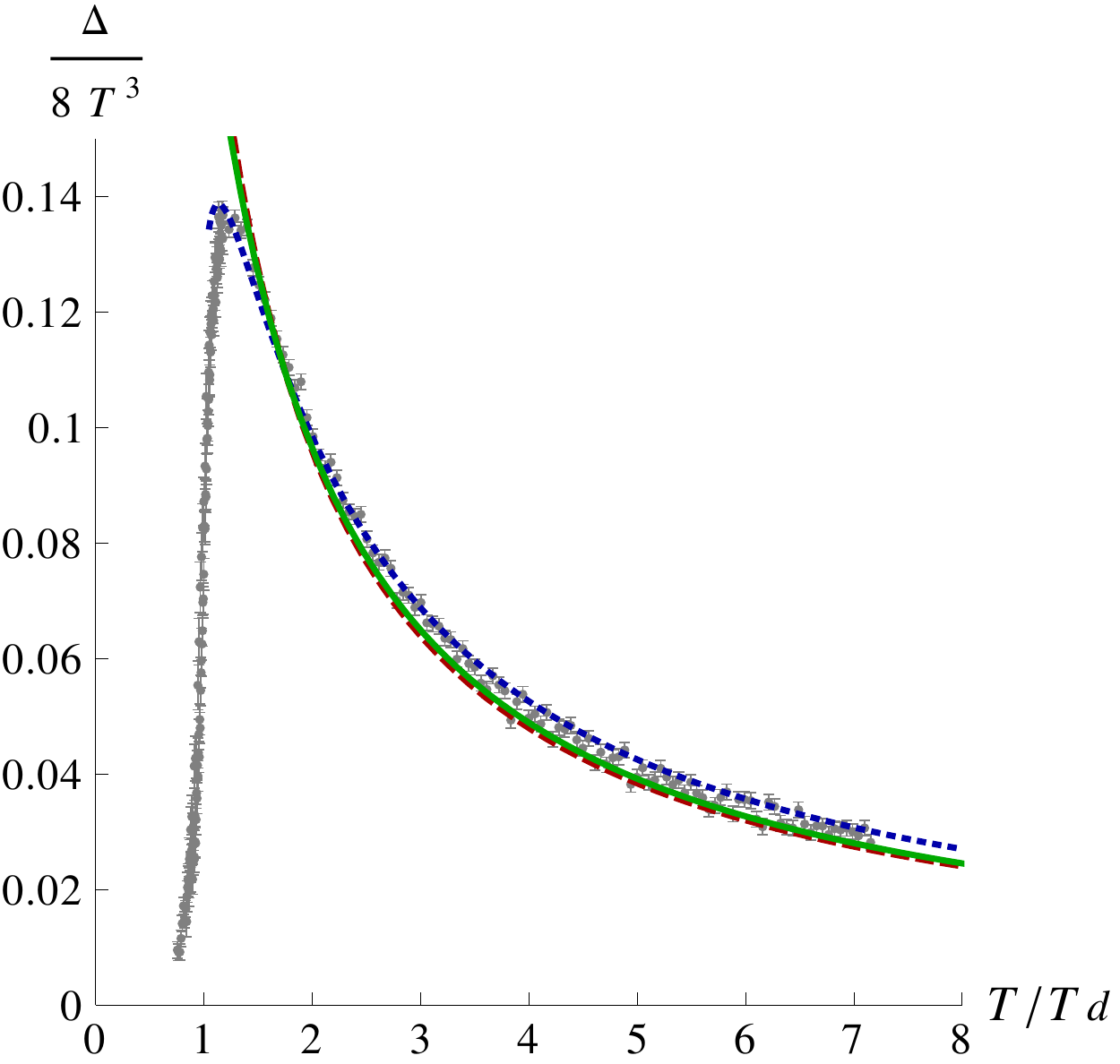}
\end{center}
\caption{Lattice interaction measure in comparison with the results for the
Vandermonde term $\sim T^{2}$. We show the curves in the one-parameter model
(dashed), the two-arameter model (solid), and in the four-parameter fit
(dotted). }%
\label{trvdm}%
\end{figure*}Notice, since $q_{min}(T)$ vanishes in the large-$T$ limit,%
\begin{equation}
\lim_{T\rightarrow\infty}{\frac{p}{3T^{3}}}\rightarrow\frac{\zeta\left(
3\right)  }{2\pi}\ =c. \label{limit}%
\end{equation}
The constant $c$ is the solution to the pressure in the perturbative limit. At
the same time, ${\frac{\Delta}{3T^{3}}}$ vanishes at large $T$.

Equations (\ref{pressureover3t3}) and (\ref{traceover3t3}) are the final
analytical results which are used to fit the lattice QCD data by adjusting the
parameters of the model.


\subsection{Vandermonde potential}

Utilizing the Vandermonde term of Eq. (\ref{vdm}), the effective potential is
given by%
\begin{align}
V_{eff}  &  =V_{pt}+V_{npt}^{B}\nonumber\\
&  =-3\frac{\zeta\left(  3\right)  }{2\pi}T^{3}\left(  1-\frac{T_{d}}{T}%
C_{3}\right) \nonumber\\
&  +T^{3}\left(  1-\frac{T_{d}}{T}C_{2}\right)  \left\{  \text{L}\left(
q\right)  -4\pi\ln\left[  2\sin(\pi q)\right]  a(T)\right\}  \text{ ,}%
\end{align}
where $V_{npt}^{B}$ is the nonperturbative contribution constructed in Eq.
(\ref{npt_B}). Repeating the analysis of Sec. \ref{sec:analyticallinear} one
can determine the minimum of the effective potential, $q_{min}(a),$ {where the
definition of the function $a(T)$ is now extended with a more general
exponent,
\begin{equation}
a=\frac{\left(  \frac{T_{d}}{T}\right)  ^{\delta}C_{1}}{\left(  1-\frac{T_{d}%
}{T}C_{2}\right)  }\text{ , \ }\delta=1,2\text{ }.
\end{equation}
The $q_{min}(a)$ we obtain for the Vandermonde potential, again with a great
accuracy, is depicted in Fig. (\ref{qmin}). At $a=a_{d}$ the linear potential
and the Vandermonde potential produce the same $V_{min}$ and $q_{min}.$}

\section{Results \label{sec:5}}

In this section we present the numerical results for the one- and the
two-parameter matrix model, and compare them to the lattice data of Ref.
\cite{Caselle:2011mn}. We show the plots for the pressure, the interaction
measure, and for the Polyakov loop utilizing three different options for the
$q$-dependent nonperturbative term $\sim C_{1}$: \ the linear term,
$T^{2}T_{d}2\pi\left[  2\ln2-\pi^{2}(q-\frac{1}{2})^{2}\right]  $, and the
Vandermonde term $T^{3-\delta}\,{T_{d}}^{\delta}4\pi\log\left[  2\sin(\pi
q)\right]  $, where we consider two different temperature dependences
$\delta=1,2.$

Close to the critical temperature the lattice data become smeared out due to
glueballs below $T_{d}$, the gluon mass above $T_{d}$, and lattice artifacts
such as finite-volume effects. Therefore, it is convenient to apply a cut, and
only fit the data at $T>1.05T_{d}$. {Moreover the finite-volume effect also
affects the pressure at high temperatures \cite{Panero:2008mg,Gliozzi:2007jh}.
In general, one finds that the pressure decreases with increasing volume.
}Motivated by the uncertainties on the lattice near $T_{d}$ and at high
temperatures, we also discuss the possibility of introducing a four-parameter
fit, and show the corresponding plots.

{We determine the free parameters of the models by applying the corresponding
nonlinear fits to the lattice pressure.} First, we present the results for the
pressure and for the interaction measure utilizing the linear term, Figs.
(\ref{plin}) and (\ref{trlin}), and the Vandermonde term, Figs. (\ref{pvdm})
and (\ref{trvdm}). For the Vandermonde term we just show the plots for the
term $\sim T^{2}$, which provides in general better fits than the other
temperature dependence $\sim T$. To give a better overview of the results we
list the values of all parameters in Table (\ref{table:parameters}).{ In this
table we also include the results of the $\chi^{2}/\mathrm{d.o.f.}$ test to
quantify how good our fit are. The lattice pressure has small error bars,
$\sim10^{-5}$. Therefore it is crucial that we achieve a high accuracy for our
ansatz for $V_{min}(a),$ ${\sim10^{-7},}$ to compare the best fits of the
different models.}\begin{table*}[t]
\begin{center}%
\begin{tabular}
[c]{c|ccccccc}\hline
Non-pert. $V$ & $C_{1}$ & $C_{2}$ & $C_{3}$ & $\delta C_{3}$ & $T_{d}$
rescale & $c$ rescale & $\chi^{2}$/dof\\\hline
1 par. vdm1 & 0.000041 & 0.999408 & 0.999940 & 0 & 1 & 1 & 265.668\\
2 par. vdm1 & 0.000000 & 1.000000 & 1.000000 & 0.024311 & 1 & 1 & 202.275\\
\hspace{2pt} 4 par. vdm1 \hspace{2pt} & \hspace{2pt} 0.003657 \hspace{2pt} &
\hspace{2pt} 0.947923 \hspace{2pt} & \hspace{2pt} 0.994911 \hspace{2pt} &
\hspace{2pt} 0.010874 \hspace{2pt} & \hspace{2pt} 0.918032 \hspace{2pt} &
\hspace{2pt} 1.031652 \hspace{2pt} & \hspace{2pt} 10.8481 \hspace{2pt}\\
1 par. vdm2 & 0.000030 & 0.999563 & 0.999956 & 0 & 1 & 1 & 285.664\\
2 par. vdm2 & 0.000000 & 0.999998 & 1.000000 & 0.025252 & 1 & 1 & 207.833\\
4 par. vdm2 & 0.006322 & 0.909976 & 0.990921 & 0.103102 & 0.855040 &
0.999717 & 1.09882\\
1 par. vlin & 0.000035 & 0.999489 & 0.999948 & 0 & 1 & 1 & 258.051\\
2 par. vlin & 0.000001 & 0.999981 & 0.999998 & 0.023831 & 1 & 1 & 199.179\\
4 par. vlin & 0.033310 & 0.525695 & 0.952861 & -0.16002 & 0.907484 &
1.014434 & 0.54232\\\hline
\end{tabular}
\end{center}
\caption{Parameters which give the best fits to the lattice pressure for
different nonperturbative terms. We use the following notation:
\textquotedblleft1 par.\textquotedblright\ for the one-parameter model,
\textquotedblleft2 par.\textquotedblright\ for the two-parameter model, and
\textquotedblleft4 par.\textquotedblright\ for the four-parameter fit.
Moreover, \textquotedblleft vlin\textquotedblright\ denotes the linear term,
\textquotedblleft vdm1\textquotedblright\ the Vandermonde term $\sim T,$ and
\textquotedblleft vdm2\textquotedblright\ the Vandermonde term $\sim T^{2}.$
The parameters $C_{2}$ and $C_{3}$ are not free, they are a function of
$C_{1}$. In \textquotedblleft1 par.\textquotedblright\ we utilize $C_{1}$ as
the single free parameter to fit the lattice data.{ In \textquotedblleft2
par.\textquotedblright\ we add a second free parameter $\delta C_{3}%
=C_{3}(T_{d})-C_{3}(\infty)$, defined in Eq. (\ref{deltac3}), to include the
effects of the bag model constant $B$. In \textquotedblleft4
par.\textquotedblright\ we further allow for small shifts in $T_{d},$ and in
the perturbative constant $c,$ in order to encompass other possible
nonperturbative effects not included in our matrix model. Moreover, we also
show the results of the $\chi^{2}/dof$ test for our fits to the lattice
pressure. }}%
\label{table:parameters}%
\end{table*}

\subsection{Results of the one-parameter model}

The one-parameter model exhibits only mild sensitivity to the choice of the
$q$-dependent nonconstant terms. By adjusting the only free parameter of the
model we already obtain good agreement with the lattice pressure and with the
interaction measure. Especially at high and low temperatures the fits are
close to the lattice data. At intermediate temperatures the agreement becomes
slightly worse. Moreover, the one-parameter model fails to reproduce the
correct shape of the peak in the interaction measure,{ residing at
$T\sim1.14T_{d}$.}

An important observation is that in the one-parameter model the best fit to
the lattice pressure gives always a rather small value of $C_{1}$, see Table
\ref{table:parameters}. From Eq. (\ref{qminT}) and Fig. (\ref{qmin}) one can
deduce that the smaller the value for $C_{1},$ the faster the condensate for
$q$ approaches zero above the critical temperature. If $q_{min}(T)\approx0$
all $q$-dependent terms in the effective potential vanish. This implies that,
except from a narrow region close to $T_{d},$ the thermodynamics is completely
governed by the $q$-independent ideal term $\sim T^{3}$ plus the constant term
$\sim T^{2}T_{d}$.

\subsection{Two-parameter model}

Aiming to improve the results of the one-parameter model, we consider the
two-parameter model,{ as proposed in Ref. \cite{Dumitru:2012fw}. }In the
two-parameter model the constants $C_{1}$ and $C_{2}$ remain the same as
before, but $C_{3}$ is replaced by the temperature-dependent parameter%
\begin{equation}
C_{3}(T)=C_{3}\left(  \infty\right)  +\frac{C_{3}\left(  T_{d}\right)
-C_{3}\left(  \infty\right)  }{T^{2}/T_{d}^{2}}\text{ ,} \label{deltac3}%
\end{equation}
which is equivalent to adding an MIT bag constant{ $B$}.

We find again that the results are quite similar for the linear term and the
Vandermonde term. The two-parameter fit improves the results of the
one-parameter model at intermediate temperatures, and gives overall good
agreement with the lattice data in the entire temperature region, see Figs.
(\ref{plin}) and (\ref{trlin}). Only at the peak of the interaction measure do
our results deviate notably from the lattice results. It must be pointed out,
however, that the parameters of the model are fixed by imposing that the
pressure vanishes at the transition point. Instead, it would be necessary to
fit the pressure in the confined phase to some hadronic (glueball) resonance
gas. Therefore, one should not expect to fit the lattice data close to $T_{d}$
with a great accuracy by making this simple assumption.

Moreover, the two-parameter model also produces an extremely narrow region in
which the condensate for $q$ is nonvanishing.

\subsection{Four-parameter fit}

The one- and the two-parameter models give already good fits to the lattice
pressure and to the interaction measure. However, at the peak of the
interaction measure, {residing close to $T_{d}$, the agreement becomes notably
worse. Further, due to the small error bars of the lattice pressure, the
$\chi^{2}/dof$ test still gives a large value $\sim$ }${200.}$

Therefore it is interesting to investigate, wether further extending the
number of degrees of freedom can improve the results near the critical
temperature,{ and reproduce the correct shape for the interaction measure}
peak. In this work, the possibility of introducing a four-parameter fit is
discussed, which can be motivated in two ways. First, in our analytical
calculations we make two obvious approximations: \ We compute the perturbative
potential only to one-loop order, and we impose that the pressure must vanish
at the transition point. Moreover, due to the smearing and finite-volume
effects present on the lattice close to $T_{d}$ and in the high-temperature
region, it is difficult to determine the exact values for the critical
temperature, and for the pressure in the perturbative limit. Taking these
uncertainties into account, two additional free parameters are introduced in
the two-parameter matrix model, one for $T_{d},$ and one for $c$, which
corresponds to the perturbative limit of the pressure, see Eq. (\ref{limit}).
We note that this four-parameter fit should be regarded just as an
approximation to a more complete model including an effective theory for the
confined phase.

The two additional parameters provide a perfect agreement with the lattice
pressure and with the interaction measure for all the three nonconstant terms
considered in this work, see Figs. (\ref{plin}), (\ref{trlin}), (\ref{pvdm}),
and (\ref{trvdm}). Especially close to $T_{d}$ the results improve notably,
giving a good fit to the peak of the interaction measure, with $\chi
^{2}/\mathrm{d.o.f.}\sim0.5$ for the model with a linear term. This shows that
the difference between the model and the lattice pressure is smaller than the
error bars.

An important result is that the four-parameter fit gives a significantly
larger value of $C_{1}$ than the other two models, see Table
(\ref{table:parameters}). This implies that there is a transition region in
the deconfined phase, in which the system develops a non-trivial condensate
for $q$, $\ q_{min}(T)\neq0$. In our matrix model this happens in principle at
all temperatures. But in practice, the condensate is only numerically large
below $\sim1.2\,T_{d}$ for the linear and for the Vandermonde term, which will
become clear when we discuss the Polyakov loop, Fig. (\ref{thlin}). This is
the range where the details of the matrix model are relevant, since the
$q$-dependent terms of the effective theory provide a nontrivial contribution
in the deconfined phase. Notably, this is in accordance with the results in
$d=3+1$, where the condensate is nonzero up to $\sim1.2T_{d}.$

In Table (\ref{table:parameters}) we list the values for the parameters. The
deviation in $c$ is rather small for all nonconstant terms, and can be
explained as follows. At high tmperature the lattice pressure is slightly
volume-dependent, and tends to decrease with increasing volume, see Ref.
\cite{Panero:2008mg,Gliozzi:2007jh}. This implies that on the lattice the
value of $c$ may be shifted to lower values when the volume is increased.
Moreover, this small shift in $c$ could be partly due to the applied one-loop
approximation. Extending the calculation to higher-loop order will shift the
perturbative constant. Thus, the higher-order loop calculations and the volume
dependence could account for the difference in $c$.

In what concerns the shift in $T_{d},$ we note that the lattice results for
the interaction measure show that there is a significant energy density below
$T_{d}$. This arises from two effects. One is simply an uncertainty of the
transition temperature, which is affected by finite-size effects such as
critical slowing down. For $N=2$ the transition is of second order. Further,
from Eq. (\ref{string_tension}), in three dimensions the ratio of $T_{d}%
/\sqrt{\sigma}$ is higher than it is in $d=3+1$, remember $\sigma$ is the
string tension. If the ratio of the glueball masses to $\sqrt{\sigma}$ is
approximately independent of the dimensionality, then the contribution of a
glueball gas to the energy density may be more significant near $T_{d}$ in
three dimensions than in four. Such effects from the confined phase are
completely neglected in our model. Ideally, we should develop an effective
theory for the confined phase, and match that to the matrix model in the
deconfined phase. Failing to do that, we adopt the prescription of the
four-parameter fit, which we admit is an approximation to a more complete theory.

\begin{figure*}[t]
\begin{center}
\includegraphics[
width=0.45\textwidth
]{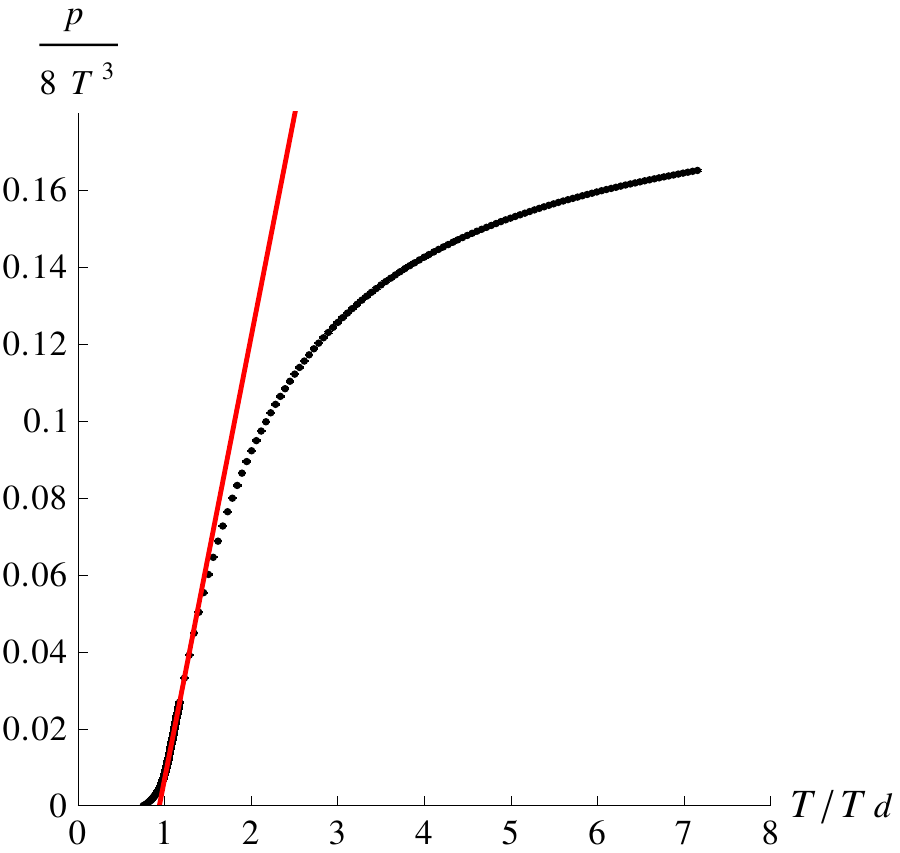}
\end{center}
\caption{We define $T_{d}$ by interpolating the tangent to the inflection
point of the pressure, which occurs at $T=1.14T_{d}$. The point where the
tangent hits the $T$-axis defines the transition temperature: $T_{d}^{\ast
}\equiv0.94.$ }%
\label{Td}%
\end{figure*}

We then define the transition temperature as the point where a linear fit to
the pressure intercepts the $T$-axis. In this case, the best estimate of
$T_{d}$ is obtained by the intercept of the tangent to the inflection point
with the $T$-axis. The inflection point is the point where the derivative is
maximum, and the second derivative vanishes. As shown in Fig. (\ref{Td}), the
intercept occurs at $0.94\,T_{d}$. This value is closer to results for the
rescaled critical temperature in the four-parameter fit, see Table
(\ref{table:parameters}).

Summing up, considering the possible systematic errors, the four-parameter fit
allows us to obtain good agreement with the lattice results in the entire
temperature range $T_{d}\leq T\leq8T_{d}$, and well reproduces the peak of the
conformal anomaly.

\subsection{Polyakov loop}

\begin{figure*}[t]
\begin{center}
\includegraphics[
width=0.45\textwidth
]{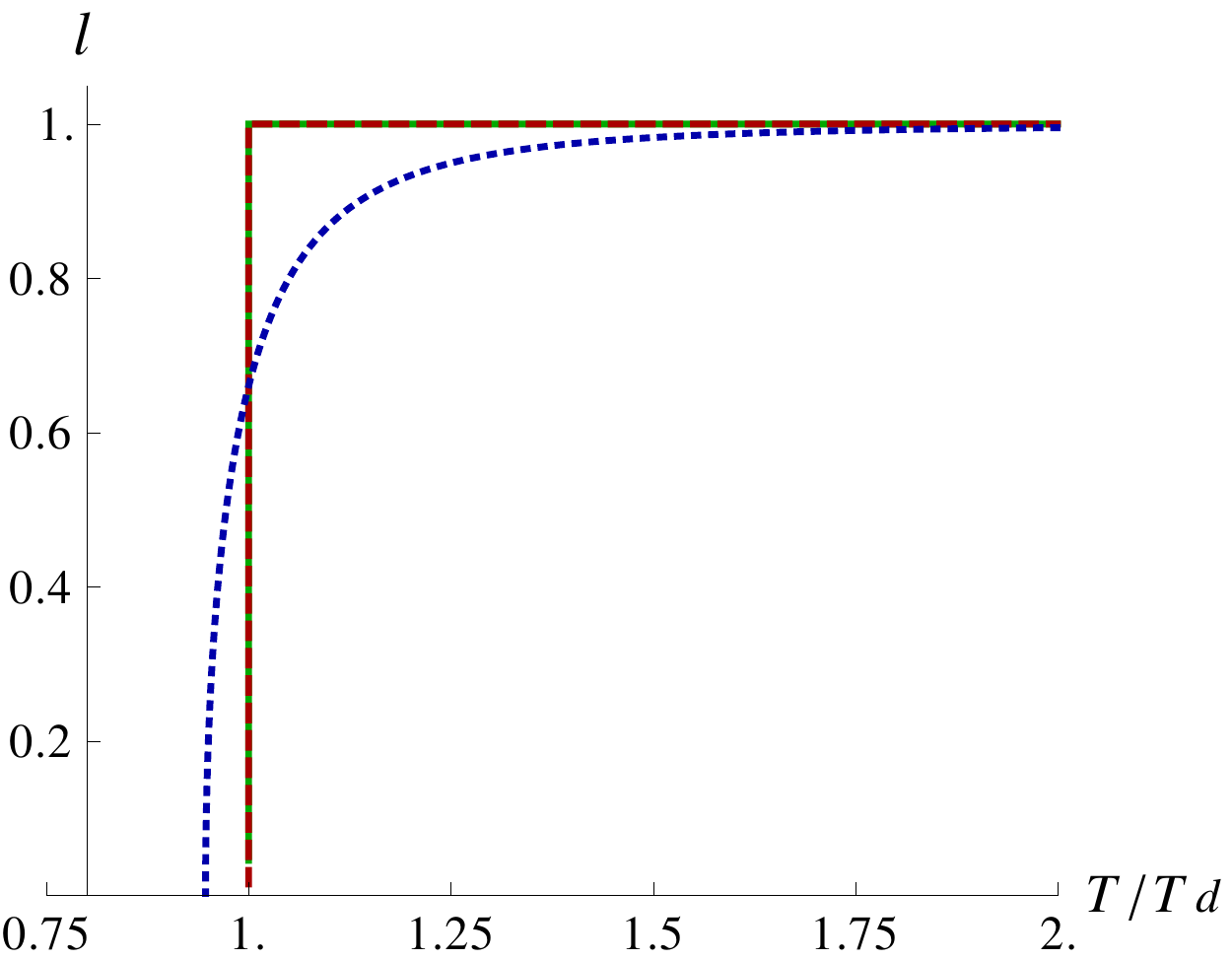} \includegraphics[
width=0.45\textwidth
]{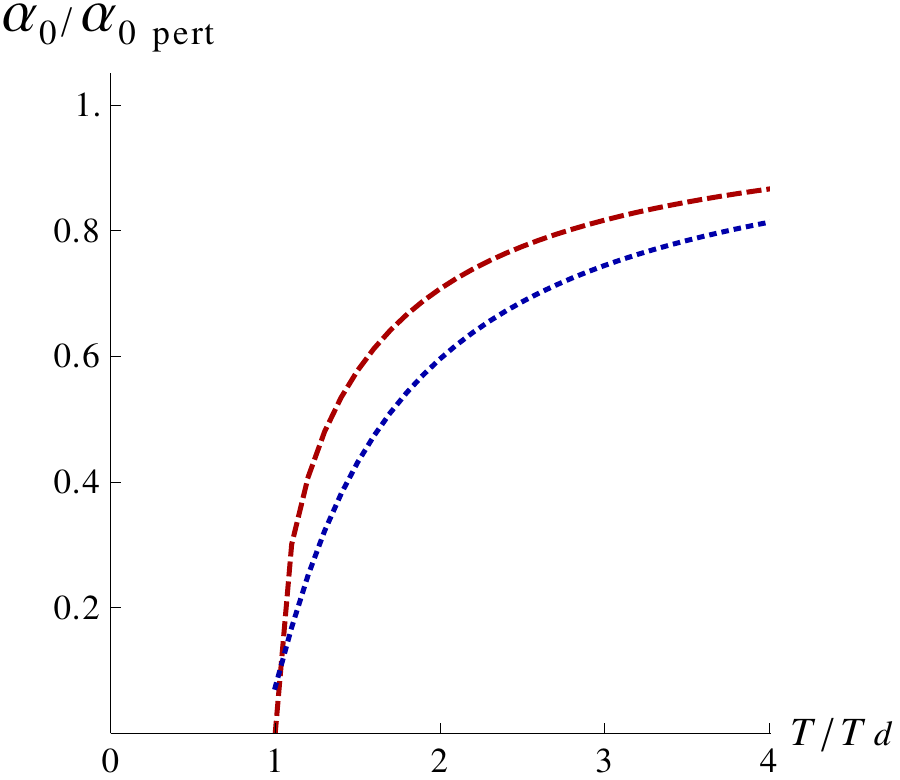}
\end{center}
\caption{Left panel: the Polyakov loop obtained using the linear term, in the
one-parameter model (dashed), the two-parameter model (solid), and in the
four-parameter fit (dotted). Right panel: the 't Hooft loop divided by its
perturbative limit, $\alpha_{0}^{pert}=5.104.$ The plots of the one- and of
the two-parameter model coincide.}%
\label{thlin}%
\end{figure*}

\begin{figure*}[t]
\begin{center}
\includegraphics[
width=0.45\textwidth
]{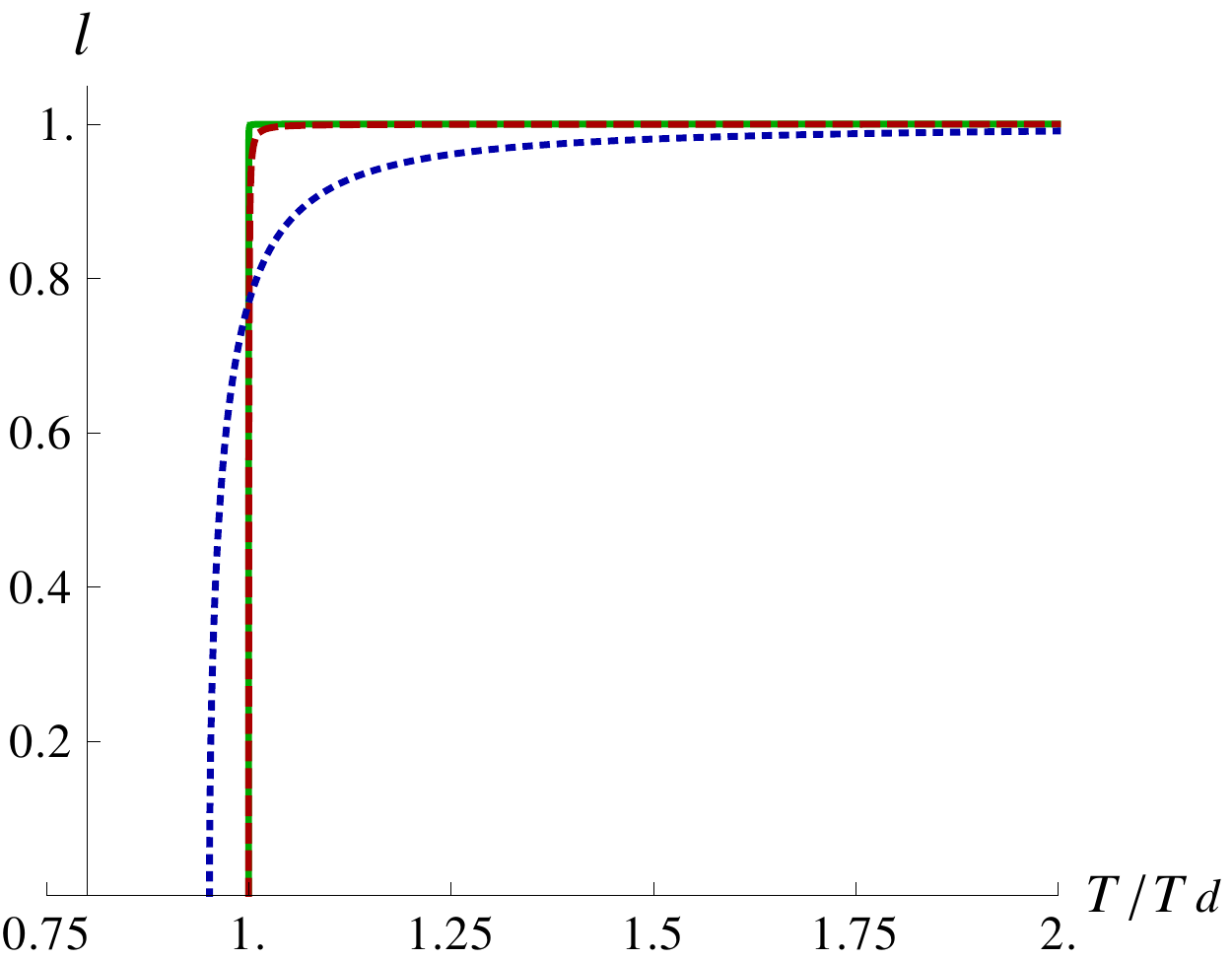} \includegraphics[
width=0.45\textwidth
]{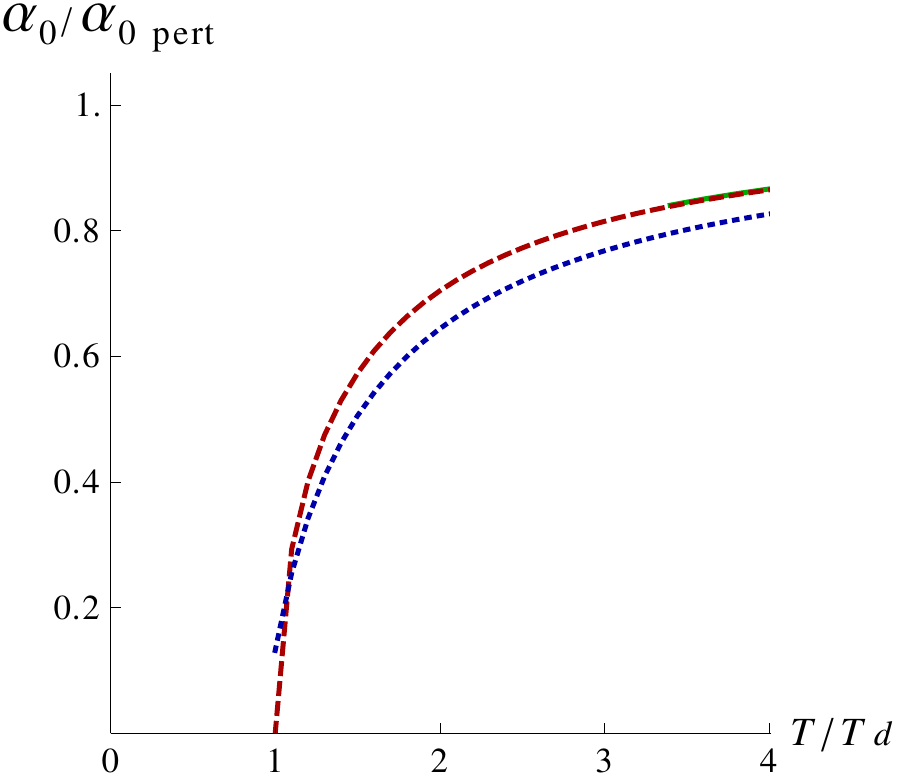}
\end{center}
\caption{The Polyakov loop (left panel), and the 't Hooft loop scaled by its
perturbative value, $\alpha_{0}^{pert}=5.104$ (right panel). The plots are
obtained utilizing the Vandermonde term $\sim T^{2},$ in the one-parameter
model (dashed), the two-parameter model (solid), and in the four-parameter fit
(dotted). For the one- and the two-parameter model the curves essentially
coincide.}%
\label{thvdm}%
\end{figure*}

Utilizing the parameters listed in Table (\ref{table:parameters}), which are
determined by fitting the lattice data for the pressure, it is possible to
compute the Polyakov loop from Eq. (\ref{p}). Figs. (\ref{thlin}) and
(\ref{thvdm}) show the Polyakov loop for the linear term and for the
Vandermonde term $\sim T^{2}$ using the one- and the two-parameter model, as
well as the four-parameter fit. In the one- and in the two-parameter model,
the Polyakov loop grows sharply from $0$ to $1$ above the critical
temperature. To understand this behavior we remember that the Polyakov loop is
given by $l\mathbf{=}\cos\left[  \pi q_{min}(T)\right]  .$ Thus, $l$\ is only
then not equal to one in the deconfined phase, if the minimum of the effective
potential differs from zero. In the one- and two-parameter model however, the
condensate, $q_{min}(T),$ is only numerically large in a narrow range close to
$T_{d},$ and then it effectively vanishes. This implies that the system merges
rapidly from confinement, $q=0,$ into the perturbative vacuum, $q=1.$

In the four-parameter fit, {which perfectly agrees with the lattice pressure,}
the condensate is non-vanishing up to $\sim1.2T_{d}$ for both nonperturbative
terms. Therefore the Polyakov loop markedly varies from one in this
temperature region. Notably, the width of the transition range is widely
independent of the details of the nonconstant terms discussed in this work.

\section{Interface tension \label{sec:6}}

In this section we construct the interface tension for our model, and present
the results for the 't Hooft loop. In absence of dynamical quarks the $SU(N)$
gauge theories exhibit a global $Z(N)$ symmetry associated with the center of
the gauge group. The confined vacuum is symmetric under $Z(N)$
transformations, whereas in the deconfined phase the $Z(N)$ symmetry is
spontaneously broken. If the system is infinite, then the spontaneous symmetry
breakdown is related to the occurrence of $N$ degenerate vacua. In a finite
volume however, bubbles of different vacua can form, which are separated by
domain walls. The dynamics of these bubbles is governed by the action of the
domain walls, which is proportional to the interface tension.

The $Z(N)$ interface tension gives the tunneling probability between two
different vacua of the system. Following the discussion of Refs.
\cite{KorthalsAltes:1996xp,Dumitru:2012fw}, we construct the interface by
putting the system in a long tube of length $2L$ in the $z$ direction, and of
length $L_{t}$ in the other two spatial directions, with $L$ $\gg$ $L_{t}\gg$
$\beta,$ and $L\rightarrow\infty$. The volume in the directions transverse to
$z$ is $\mathcal{V}_{tr}=\beta L_{t}$.\ To model the interface tension we
assume that the system is in a vacuum state at both ends, but not in between.
This forces a $Z(N)$ interface along the $z$-direction. The action of the
interface is equal to the interface tension $\alpha$, times the transverse
volume, $\mathcal{V}_{tr}$%
\begin{equation}
\alpha=\frac{S}{\mathcal{V}_{tr}}\text{ .}%
\end{equation}
To compute the interface tension one first needs to construct the effective
action $S,$ which is given by the effective potential plus a kinetic term%
\begin{equation}
S=\mathcal{V}_{tr}\int dz\left[  \mathcal{T}_{kin}(q)+V_{eff}(q)\right]
\text{ .}%
\end{equation}
At leading order, for $q$ varying slowly on the scale of $1/T,$ it is
sufficient to use the kinetic term at tree level, which is given by the
classical action%
\begin{align}
\mathcal{T}_{kin}(q)  &  =\frac{1}{2}\operatorname{tr}G_{\mu\nu}^{2}=\frac
{\pi^{2}T^{2}}{g^{2}}\left(  \frac{dq}{dz}\right)  ^{2}\operatorname{tr}%
\sigma_{3}^{2}\label{cl}\\
&  \equiv\frac{T^{3}}{2}\left(  \frac{dq}{dz\prime}\right)  ^{2}\text{
,}\nonumber
\end{align}
where we introduce the rescaled coordinate $z$%
\begin{equation}
\text{ }z\prime=\frac{z}{\gamma}\text{ , \ \ }\gamma=\frac{2\pi}{g\sqrt{T}%
}\text{ .} \label{z}%
\end{equation}
$\gamma$ is the parameter which controls the width of the domain wall between
the two vacua. Notice, at the classical level the action reduces to only a
kinetic term, since the classical field of Eq. (\ref{a0}) commutes with
itself. This means that classically there is no difference between the two vacua.

Assuming that the vacua at the two ends of the box, $z=-L$ and $z=+L$,
correspond to the two minima, $q_{i}$ and $q_{f}$, the interface tension is
connected to the shortest path between $q_{i}$ and $q_{f}$ , which obeys the
equation of motion%
\begin{equation}
T^{3}\frac{d^{2}q}{dz\prime^{2}}=\frac{dV_{eff}(q)}{dq}\text{ }, \label{eqm}%
\end{equation}
with the boundary conditions $q(-L)=q_{i}$ and $q(L)=q_{f}$. The corresponding
energy density is obtained by multiplying Eq. (\ref{eqm}) by $dq/dz\prime$,
and integrating over $z\prime$%
\begin{equation}
e=\frac{T^{3}}{2}\left(  \frac{dq}{dz\prime}\right)  ^{2}-V_{eff}(q).
\end{equation}
For any solution to the equation of motion the energy is conserved,
$de/dz\prime=0$. Therefore, \bigskip any function $q(z)$ which minimizes the
effective action with respect to the corresponding boundary conditions
satisfies:%
\begin{equation}
\text{ }\frac{T^{3}}{2}\left(  \frac{dq}{dz\prime}\right)  ^{2}=V_{eff}%
(q)\text{ , \ \ \ }\frac{dq}{dz\prime}=\sqrt{\frac{2V_{eff}(q)}{T^{3}}}\text{
.} \label{ec}%
\end{equation}
Using the energy conservation in Eq. (\ref{ec}), the effective action can be
written as%
\begin{align}
S  &  =2\mathcal{V}_{tr}\int dzV_{eff}(q)\nonumber\\
&  =\gamma\sqrt{T^{3}}\mathcal{V}_{tr}\int_{q_{i}}^{q_{f}}dq\sqrt{2V_{eff}%
(q)}\nonumber\\
&  =\frac{2\pi}{g}T\mathcal{V}_{tr}\int_{q_{i}}^{q_{f}}dq\sqrt{2V_{eff}%
(q)\text{ }}\text{ .}%
\end{align}
The general form for the interface tension\ is then%
\begin{equation}
\alpha=\alpha_{0}\frac{\sqrt{T^{5}}}{g}\text{ , }%
\end{equation}
where we define the dimensionless quantity%
\begin{equation}
\alpha_{0}=2\pi\int_{q_{i}}^{q_{f}}dq\sqrt{\frac{2V_{eff}(q)}{T^{3}}}\text{ }.
\end{equation}
It is interesting to notice that the factor $1/g^{2}$ present at the classical
level {in Eq. (\ref{cl}), becomes $1/g.$} This is because the effective action
acquires a potential only at one-loop order. Furthermore, from the definition
of the rescaled length $z\prime$ in Eq. (\ref{z}) follows that the relevant
distance scale in the effective action is not $1/T$ , but $1/g\sqrt{T}$.
Therefore, if the coupling constant is small, the effective action varies over
much larger distance scales than $1/T$ . This implies that in weak coupling
the variation of $q(z)$ in space is slow and can be ignored.

\subsection{The order-order interface tension}

Above the deconfinement temperature the theory can be in one $Z(N)$ vacuum,
$q_{i}=q_{min}^{1}(T),$ at one end of the box, and in a degenerate but
inequivalent vacuum, $q_{f}=q_{min}^{2}(T),$ at the other end. Due to the
$Z(2)$ symmetry $q_{min}^{2}(T)=1-q_{min}^{1}(T).$ This is the order-order
interface tension, which is equivalent to a 't Hooft loop in the deconfined
phase. The associated tunneling probability is determined by the integral%
\begin{equation}
\alpha_{0}=2\pi\int_{q_{min}^{1}(T)}^{q_{min}^{2}(T)}dq\sqrt{\frac
{2V_{eff}(q)}{T^{3}}}\text{ ,}%
\end{equation}
where $V(q)$ is the difference between the effective potential in $q$ and at
the minimum%
\begin{equation}
V(q)=V_{eff}(q,T)-V_{eff}\left[  q_{min}^{1}(T)\right]  \text{ .}%
\end{equation}
Figures \ref{thlin} and \ref{thvdm}\ show the plots for the 't Hooft loop
scaled by its perturbative limit, $\alpha_{o}^{pert}=5.104.$ In the
one-parameter model the results are essentially the same when using the linear
term and the Vandermonde term. One can understand this by remembering that in
the one-parameter model the minimum of the effective potential merges rapidly
from the confined vacuum, $q_{min}(T_{d})=q_{c}=0.5$, into the perturbative
vacuum, $q_{min}(T>T_{d})\approx0.$ Therefore, the two degenerate minima are
approximately at $q_{min}^{1}(T)\approx0$ and $q_{min}^{2}(T)\approx1.$
Remarkably, unlike the Polyakov loop, which becomes trivial in the
one-parameter model, for the 't Hooft loop the details of the matrix model are
relevant in the entire semi-QGP for all models addressed in this work.

\section{Conclusions and outlook \label{sec:7}}

In this work we utilize a matrix model to study the deconfinement phase
transition in pure $SU(2)$ glue theory in $2+1$ dimensions. The basic
{variables of the model are the eigenvalues} of the Wilson line. First we
construct the effective potential as the sum of a perturbative and a
nonperturbative part. The perturbative potential is computed in the presence
of a constant background field for the vector potential $A_{0}\sim q.$ We find
that to one-loop order this gives a trilogarithm function of $A_{0}/T.$ Then,
in order to model the transition to deconfinement, we introduce additional
constant and nonconstant nonperturbative terms depending on $T,$ and on three
parameters. For the nonconstant terms, which are functions of $q,$ we try
three different Ans\"{a}tze: the linear term $\sim T^{2}T_{d}2\pi\left[
2\ln2-\pi^{2}(q-\frac{1}{2})^{2}\right]  $, and a Vandermonde-like term with
two different temperature dependences, $\sim T^{3-\delta}\,{T_{d}}^{\delta
}4\pi\log\left[  2\sin(\pi q)\right]  ,$ $\delta=1,2$. Imposing two
constraints for the phase transition at $T=T_{d}$ leaves only one free
parameter, which is determined by fitting the lattice pressure. The numerical
results for the pressure and for the interaction measure are presented and
compared to the {lattice data of Ref. \cite{Caselle:2011mn}.}

The one-parameter model already gives{ good} fits to the lattice pressure and
to the interaction measure at high and low temperatures. But at intermediate
temperatures the results deviate from the lattice results. The two-parameter
model improves the agreement at intermediate temperatures, and provides
overall{ good} fits to the pressure, and to the interaction measure at all
temperatures. However, in both models there is a clear deviation from the
lattice data at the peak of the interaction measure, {while their $\chi^{2}%
$/\textrm{d.o.f.} tests indicate that better fits are possible}. Considering
different options to cure this deficiency, the possibility of constructing a
four-parameter fit is discussed. Regarding possible uncertainties present in
our analytical calculations, due to the applied approximations, as well as on
the lattice, due to glueballs and finite-volume effects, the two-parameter
model is extended by two additional free parameters: one for $T_{d},$ and one
for the perturbative limit of the pressure, $c.$ The four-parameter model
gives remarkably good fits to the lattice pressure and to the interaction
measure for all nonconstant terms discussed in this work. It also reproduces
the correct shape for the peak of the conformal anomaly. Furthermore, {in the
four-parameter fit there is a range in the deconfined phase, where the
condensate is nonzero, and the details of the matrix model become relevant.
The window of this transition region extends up to $\sim1.2$ $T_{d}.$ This is
similar to the results for the }$SU(2)${ matrix model obtained }in $d=3+1$%
{$.$} We remark however, that this four-parameter fit should be considered
just as a possible approximation to a more complete model which involves an
underlying effective theory for the confined phase. Notably, the one- and the
two-parameter model, as well the four-parameter fit exhibit only a mild
sensitivity to the details of the nonconstant terms.

Using the parameters determined by fitting the pressure, we also show the
plots for the Polyakov loop and for the 't Hooft loop. In the one- and in the
two-parameter model the Polyakov loop grows sharply from $0$ to $1$ above
above the critical temperature. This is because in our model the Polyakov loop
differs from one only when the condensate for $q$ is nonvanishing. In the one-
and the two-parameter model, however, the condensate effectively vanishes
rapidly above $T_{d}.$ In the four-parameter fit{ the transition range where
the condensate is nonvanishing and where the Polyakov loop varies from one
extends up to $\sim1.2\;T_{d}.$}

The model can be improved in two obvious ways. First, one can include
perturbative corrections at next to leading order, to $\sim g^{2}$. This will
presumably correct the deviation from the lattice data at high temperature.
Second, near $T_{d}$ it is necessary to include an effective theory for the
confined phase. This will describe the increase in the energy density near
$T_{d}$, and obviate our rather ad hoc prescription for shifting $T_{d}$ by hand.

Summing up, the one- and the two-parameter matrix models work reasonably well
for the pressure and for the interaction measure. They also provide reasonable
predictions for the 't Hooft loop. The four-parameter fit agrees perfectly
with the lattice data even very close to $T_{d}$. Moreover, it provides
reasonable results\ for the Polyakov loop. This is closely related to the
width of the transition region, in which the model exhibits a nontrivial
minimum. So far, the behavior of the Polyakov loop and of the 't Hooft near
$T_{d}$ in $d=2+1$ have not been computed on the lattice. These results could
provide important tests of our model.

\begin{acknowledgments}
The authors would like to thank Marco Panero
for kindly sharing the lattice data of Ref.
\cite{Caselle:2011mn}. We also thank Dirk H. Rischke, Nuno Cardoso and Marco 
Panero for valuable discussions. The research of R.D.P. is supported
by the U.S. Department of Energy under contract \#DE-AC02-98CH10886.
E. S. thanks the hospitality of RIKEN/BNL and CFTP.
The research of P.B. is supported by the CFTP grant
PEST-OE/FIS/UI0777/2011, the FCT grant  CERN/FP/123612/2011, and the
CRUP/DAAD exchange A10/10.
\end{acknowledgments}

\bibliographystyle{apsrev4-1}
\bibliography{matrix_model}

\end{document}